\documentclass{sig-alternate-10pt}

\usepackage{balance}  %
\usepackage{graphics} %
\usepackage{times}    %
\usepackage{url}      %

\usepackage{verbatim} %

\usepackage{color}

\newcommand{\textred}[1]{\textcolor{red}{#1}}
\ifx\noeditingmarks\undefined
   \newcommand{\pgwrapper}[2]{\textred{#1 #2}}
\else
   \newcommand{\pgwrapper}[2]{}
\fi

\makeatletter
\def\url@leostyle{%
  \@ifundefined{select}{\def\UrlFont{\sf}}{\def\UrlFont{\small\bf\ttfamily}}}
\makeatother
\def\pprw{8.5in}
\def\pprh{11in}

\usepackage[pdftex]{hyperref}
\hypersetup{
pdftitle={User-Generated Free-Form Gestures for Authentication: Security and Memorability},
pdfauthor={Janne Lindqvist},
pdfkeywords={free-form gestures, security, mobile authentication},
bookmarksnumbered,
pdfstartview={FitH},
colorlinks,
citecolor=black,
filecolor=black,
linkcolor=black,
urlcolor=black,
breaklinks=true,
}

\begin{document}

\title{User-Generated Free-Form Gestures for Authentication: Security and Memorability}

\numberofauthors{1}
\author{
  \alignauthor Michael Sherman$^\dagger$, Gradeigh Clark$^\dagger$, Yulong Yang$^\dagger$, Shridatt Sugrim$^\dagger$, Arttu Modig$^\ast$, Janne Lindqvist$^\dagger$, Antti Oulasvirta$^\ddagger$, Teemu Roos$^\ast$ \\
  $^\dagger$Rutgers University, $^\ddagger$Max-Planck Institute for Informatics, $^\ast$University of Helsinki \\ Corresponding author: janne$@$winlab.rutgers.edu \\
  }

\maketitle

\begin{abstract}
This paper studies the security and memorability of free-form multitouch gestures for mobile authentication. Towards this end, we collected a dataset with a generate-test-retest paradigm where participants (N=63) generated free-form gestures, repeated them, and were later retested for memory. Half of the participants decided to generate one-finger gestures, and the other half generated multi-finger gestures. Although there has been recent work on template-based gestures, there are yet no metrics to analyze security of either template or free-form gestures. For example, entropy-based metrics used for text-based passwords are not suitable for capturing the security and memorability of free-form gestures. Hence, we modify a recently proposed metric for analyzing information capacity of continuous full-body movements for this purpose.  Our metric computed estimated mutual information in repeated sets of gestures. Surprisingly, one-finger gestures had higher average mutual information. Gestures with many hard angles and turns had the highest mutual information. The best-remembered gestures included signatures and simple angular shapes.  We also implemented a multitouch recognizer to evaluate the practicality of free-form gestures in a real authentication system and how they perform against shoulder surfing attacks. We conclude the paper with strategies for generating secure and memorable free-form gestures, which present a robust method for mobile authentication.
\end{abstract}

\section{Introduction}

Smartphones and tablets today are important for secure daily transactions. They are part of multi-factor authentication for enterprises \cite{azuremultifactor}, allow us to access our email, make 1-click payments on Amazon, allow mobile payments \cite{square} and even access to our houses \cite{gogogate}. Therefore, it is important to ensure the security of mobile devices. 

Recently, mobile devices with touchscreens have made gesture-based authentication common. For example, the Android platform includes a 3x3 grid that is used as a standard authentication method, which allows users to unlock their devices by connecting dots in the grid. Compared with text-based passwords, gestures could be performed faster while require less accuracy.
Although grid-based gestures better utilize the capabilities of touchscreens as input devices, they are limited as an authentication method. For example, a visual pattern drawn on a grid is prone to attacks such as shoulder surfing~\cite{Zakaria:2011:SSD:2078827.2078835} and smudge attacks~\cite{Aviv:2010:SAS:1925004.1925009}.

\begin{figure}[t!]
\centering
\includegraphics[scale=0.598]{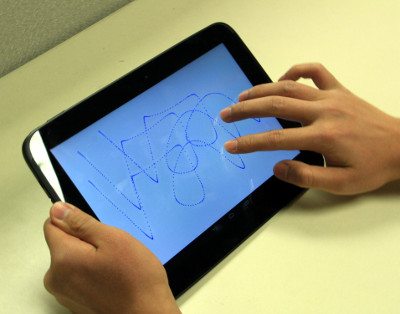}
\caption{This paper studies continuous free-form multitouch gestures as means of authentication on touchscreen devices. Authentication on touchscreens is normally done with a grid-based method. Free-form gesture passwords have a larger password space and are possibly less vulnerable to shoulder surfing. We note that there are no visual cues for the gestures, the gesture traces are shown only after creating the gesture.}
\label{fig:example_photo}
\end{figure}

This paper studies \emph{free-form multitouch gestures} without visual reference; that is, gestures that allow all fingers draw a trajectory on a blank screen with no grid or other template. An example of the
creation process is depicted in Figure~\ref{fig:example_photo}, where the gesture traces are shown only after the gesture was created.
This method bears potential, because it relaxes some of the assumptions that make the grid-based methods vulnerable. 
In particular, arbitrary shapes can be created.
Moreover, as more fingers can be used,  in principle more information can be expressed.
Technically such gestures can be scale and position invariant, allowing the user to gesture on the surface without visually attending the display. Consider, for example, drawing a circle as your password.
This may be beneficial for mobile users who need to attend their environment.
Nevertheless, although no visual reference is provided, mnemonic cues referring to shapes and patterns can still be utilized for generating the gestures. 
Finally, when multiple fingers are allowed to move on the surface and no visual reference is provided,  observational attacks may be more difficult.

Previous work on gestures as an authentication method has focused on two major directions: one was whether the same gesture can be correctly recognized in general \cite{touchalytics,Sae-Bae:2012:BGN:2207676.2208543} or in  a specific environment such as handwriting motion detected by Kinect-cameras \cite{kinwrite}, predefined whole-body gestures detected from wireless signals \cite{Pu:2013:WGR:2500423.2500436}, and mobile device movement detected by built-in sensors \cite{Ruiz:2011:UMG:1978942.1978971}. Studies of the security of gestures look at either the protection of gestures from specific scenarios \cite{Zakaria:2011:SSD:2078827.2078835,Schaub:2013:EDS:2501604.2501615,kinwrite,DeLuca:2013:BAS:2470654.2481330}, or an indirect measurement of security \cite{Grandhi:2011:UNI:1978942.1979061,Oh:2013:CPE:2470654.2466145}. Further, these works have focused on understanding performance of template gestures repeated by participants, not user-generated free-form gestures as the present work. %

Our goal is to understand the security of this method by measuring information capacity and studying memorability in a dataset that allowed users to freely choose the kind of multitouch passwords they deemed best.
We conducted a controlled experiment with 63 participants in a generate-test-retest design. 
At first, participants created and repeated gestures (generate), then trying to recall it after a short break (test) and recalling it again after a period of time at least 10 days (retest). With this paradigm we were able to examine the effect of time on how participants memorize their gestures. To the best of our knowledge, we are the first to present a study how people actually recall their gestures after a delay. 

To analyze the security of the gestures, we use a novel information metric of mutual information in repeated multifinger trajectories.  We base our metric on a recent metric that was used for a very different purpose, specifically the estimation of throughput (bits/s) in continuous full-body motion \cite{oulasvirta2013information}, and it has not been used previously for authentication.
Because multitouch gestures are continuous by nature the standard information metrics cannot be directly applied. What is unique to gesturing over discrete aimed movements (physical and virtual buttons) is that every repetition of a trajectory is inherently somewhat different \cite{jones2006human}. However, when this variability grows too large, the password is useless, because it is both not repeatable by the user and not discriminable from other passwords. 
The information metric should capture this variability. 
In our metric, a secure gesture should contain a certain amount of ``surprise", i.e. some turns or changes, while still being able to be reproduced by the user itself. We also include mutual information calculation to separate the complexity of controlled and intended features of the gesture and that of uncontrolled and unreproducible features. 

Our results show that several participants were able to create secure and memorable gestures without guidance and prior practice. However, many participants
used multiple fingers in a trivial way, just repeating the same gesture. Our implementation of a practical multitouch recognizer shows that the
free-form gestures can used as a secure authentication mechanism, and are resistant to shoulder-surfing attacks.

Our contributions are as follows:
\begin{enumerate}
\item Report on patterns in user-generated free-form multitouch gestures generated from 63 participants with a typical tablet;
\item Adaptation of a recent information theoretic metric for measuring the security and memorability of gestures;
\item A design and implementation of a practical multitouch gesture recognizer to evaluate free-form gestures applicability for authentication;
\item A preliminary study on a shoulder-surfing attack that indicates the potential of free-form gestures against such attacks. 
\end{enumerate}

\section{Related Work}

In this section, we discuss related work on biometric-rich authentication schemes, graphical passwords, and password memorability.

\textbf{2D gesture authentication schemes.}
Similar to free-form gestures, biometric-rich authentication schemes are based on the premise that when a user performs a gesture on a touchscreen they will do this in such a way that features can be extracted that will uniquely identify them later on \cite{touchalytics,Sae-Bae:2012:BGN:2207676.2208543, WM-CS-2012-06,Bo:2013:SSU:2500423.2504572}. Similar ideas have been applied to recognizing motions with Kinect~\cite{kinwrite}. Specifically, Sae-Bae et al.~\cite{Sae-Bae:2012:BGN:2207676.2208543} has shown that there is a uniqueness to the way users perform identical set of template 2D gestures based on biometric features (e.g.~hand size and finger length). Frank et al.~\cite{touchalytics} demonstrated that the way a user interacts with a smartphone forms a unique identifier for that user; they show that the way a user performs simple tasks (e.g.~scrolling to read or swiping to the next page)  is performed in a unique way such that the coordinates of a stroke, time, finger pressure, and the screen area covered by a finger are measurements that could be used to classify said user. Zheng et al.~\cite{WM-CS-2012-06}, operating on similar principles, have studied behavioral authentication using the way a user touches the phone -- the features extracted included acceleration, pressure, size, and time. Bo et al.~\cite{WM-CS-2012-06} performed recognition by mining coordinates, duration, pressure, vibration, and rotation.  Cai et al.~\cite{book} examined six different features (e.g.~sliding) and compared data such as the speed, sliding offset, and variance between finger pressures. Deluca et al.~\cite{DeLuca:2013:BAS:2470654.2481330} developed a system for authentication by drawing a template 2D gesture on the back of a device using two phones connected back to back. %
The security of the gesture is analyzed through various methods by an attacker to replicate the original biometric or graphical password -- there is no analysis performed as to the security content of the gesture, just its difficulty to be reproduced. 

\textbf{3D gesture authentication schemes.} 3D gesture recognition can be performed, most recently, using camera-based systems (e.g.~Kinect)~\cite{kinwrite} or using wireless signals \cite{Pu:2013:WGR:2500423.2500436}. With the camera-based systems, a user would trace a gesture out in space and the image gets compressed into a two dimensional image and processed for recognition \cite{kinwrite}.  Pu et al.~\cite{Pu:2013:WGR:2500423.2500436} have shown that three-dimensional gestures can be recognized by measuring the Doppler shifts between transmitted and received Wi-Fi signals.  %

\textbf{Graphical and text-based passwords security and memorability.} There has been considerable work on cued graphical passwords, a survey is offered by Biddle et al.~\cite{Biddle:2012:GPL:2333112.2333114} for the past twelve years.
In particular, there has been analysis on how Draw a Secret (DAS) \cite{Jermyn:1999:DAG:1251421.1251422} type of graphical passwords measures up to text-based passwords in terms of dictionary attacks~\cite{Oorschot:2008:PMU:1284680.1284685}. Oorschot et al.~\cite{Oorschot:2008:PMU:1284680.1284685} go on to describe a set of complexity properties based for DAS passwords  and conclude that symmetry and stroke-count are key in how complicated a DAS-password can be. They do not provide a direct measurement of this for DAS-password; the analysis is restricted to constructing a model to perform a dictionary attack and show that there are weak password subspaces based on DAS symmetry. For text-based passwords Florencio et al.~\cite{Florencio:2007:LSW:1242572.1242661} studied people's web password habits, and found that people's passwords were generally of poor quality, they are re-used and forgotten a lot.  Yan et al.~\cite{Yan:2004:PMS:1024867.1025014} were among the first to study empirically how different password policies affect security and memorability of the text-based passwords. Chiasson et al.~\cite{Chiasson:2009:MPI:1653662.1653722} conducted laboratory studies on how people recall multiple text-based passwords compared to multiple click-based graphical passwords (PassPoints \cite{Wiedenbeck:2005:PDL:1090412.1090418}). They found that the recall rates after two weeks were not statistically significant from each other.  Everitt et al.~\cite{Everitt:2009:CSF:1518701.1518837} analyzed the memorability of multiple graphical passwords (PassFaces \cite{Biddle:2012:GPL:2333112.2333114}) through a longitudinal study and found that users who authenticate with multiple different graphical passwords per week were more likely to fail authentication than users who dealt with just one password. 

\textbf{Security Analysis of Graphical Passwords and Gestures.} Most security analysis focus on preventing shoulder surfing attacks from hijacking a graphical password or gesture \cite{Zakaria:2011:SSD:2078827.2078835,Schaub:2013:EDS:2501604.2501615,DeLuca:2013:BAS:2470654.2481330}. The methods depend on implementing techniques to make the input more difficult to attack (e.g.~making the graphical password disappear as it is being drawn \cite{Zakaria:2011:SSD:2078827.2078835}). Another team designed an algorithm based on Rubine~\cite{Rubine:1991:SGE:122718.122753} that told users whether or not their gestures are too similar, although the metric for this is inherently based on the recognizer's scoring capabilities and not on a measure of the gesture by itself \cite{Long:2001:LSI:971478.971510}.  Schaub et al.~\cite{Schaub:2013:EDS:2501604.2501615} suggest that the size of the password space for a gesture is based on three spaces: design features (how the user interacts with the device), smartphone capabilities (screen size, etc.), and password characteristics (existing metrics of security, usability, etc). Security in this context refers to a measured resistance to shoulder surfing. 

Continuing with security analysis, brute force attacks on gestures have been examined in some studies \cite{kinwrite,zhao:picturegesture,Biddle:2012:GPL:2333112.2333114}. Zhao et al.~\cite{zhao:picturegesture} have examined the security of 2D gestures against brute force attacks (assisted or otherwise) when using an authentication system where a user will draw a gesture on a picture. A measure of the password space is developed and an algorithm under which a gesture in that space can be attacked. The attack is capable of guessing the password based on areas of the screen that a user would be drawn towards. This study does not concern itself with the security of the gesture drawn, instead it is focused on where a user would target in a picture-based authentication schema -- it does not address free-form gesture authentication. Serwadda et al.~\cite{Serwadda:2013:KTB:2508859.2516659} showed that authentication schema based on biometric analysis (including one by Frank et al.~\cite{touchalytics}) can be cracked using a robot to brute force the inputs using an algorithm that is supplied swipe input statistics from the general population. 

Finally, on non-security related work, Oulasvirta et al. \cite{oulasvirta2013information} studied the information capacity of continuous full-body movements. Our metric is motivated by their work. Specifically, they did not study 2D gestures or their security and memorability or use for an authentication system. When asked to create gestures for non-security purposes, previous work \cite{Grandhi:2011:UNI:1978942.1979061, Oh:2013:CPE:2470654.2466145} indicates that  people tend to repeat gestures that are seen on a daily basis and are context-dependent (e.g. that the gestures people perform are dependent on whether they are directing someone to perform a task or receiving directions on a task).

\section{Security of Gestures}

In this section, we present our novel information-theoretic metric for evaluating the security and memorability of gestures. 
We briefly discuss why existing entropy-based metrics used to evaluate discrete text-based passwords \cite{Bonneau:2012:SGA:2310656.2310721} are
not suitable for gestures, and move to present our metric for security and memorability of continuous gestures. We have modified a recent
metric on analyzing information capacity of full-body movements \cite{oulasvirta2013information} to estimate the security of a multitouch gesture.

Multitouch gestures on a touchscreen surface produce trajectory data where the positions of one or more end-effectors (finger tips) are tracked over time.
The continuous and multi-dimensional nature of multitouch gesture data poses some challenges for defining the information content
compared to regular keyboard-based passwords that only gauge information in \emph{discrete movements} corresponding to key events (pressdown) caused by a single end-effector (e.g., finger, cursor) at a time.
\emph{Multitouch gestures} involve multiple end-effectors and continuous movement.
To our knowledge, no information theory based security metric has been proposed for multitouch gestures as passwords.

The core idea is to demonstrate that there is an association between the security of a gesture password and the \emph{information content} of the gesture. Intuitively, information content is a property of a
message or a signal (such as a recorded gesture): it measures
the amount of surprisingness, or unpredictability, of the signal
with the important additional constraint that any surprisingness
due to random (uncontrolled) component in the signal is
excluded. Information-theoretically, the surprisingness of a message, or more precisely, of a source generating messages according to a certain probability distribution, can be measured by the entropy $H(x)$ associated with the random variable, $x$, whose values are the messages. For instance, the surprisingness of a key stroke chosen uniformly at random among 32 alternatives is $\log_2(32) = 5$ 
bits; five times that of an answer to a single yes--no question.
A similar measure of surprisingness can be also associated to continuous random variables, called the differential entropy, but it lacks a similar meaning in terms of yes--no questions as it can, for instance, take negative values. For more detailed definitions of the used information-theoretic concepts, please see e.g. Cover and Thomas~\cite{cover-thomas}. 

For discrete as well as continuous messages, the information content can be defined as the \emph{mutual information} $I(x;y)$ between two random variables, $x$ and $y$, and it gives the reduction in the entropy (surprisingness) of one random variable ($y$) when another one ($x$) becomes known. Note that a message can have high entropy (complexity) without the mutual information (information content) being high but not vice versa.\footnote{In fact, to be precise, the inequality $I(x;y) \leq \min\{H(x),H(y)\}$ holds only for discrete signals, such as symbolic passwords, but not for continuous signals, such as gestures, because continuous signals can have negative entropy~\cite{cover-thomas}. However, even though there is no theoretical guarantee of it, the intuition that a trivial gesture such as a straight line cannot contain high information content holds true in our experiments; see below.} 

The metric we use for the information content in repeated gestures
is defined as the mutual information between two realizations of the
same gesture. The underlying assumption is that any dependencies
between the two gestures must be intended, and conversely, any aspects of the gestures that are not present in both realizations must be unintended. 

The input to the metric comprises of two multitouch movement sequences produced by asking a user to produce a gesture and repeat it.
The two gesture trajectories are denoted by $x$ and $y$.
The trajectories record the locations of each of the used fingers
over duration of the gesture. 
The information content is then the amount of information, as
outlined above in the information-theoretic sense, that is contained in both $x$ and $y$, measured by the mutual information.

\textbf{Computation.} Computation of the mutual information involves a sequence of steps.
First, we need to remove from the sequences their predictable
aspects, as far as possible. To do so, we fit a second order autoregressive model for both of the sequences separately. 
For sequence $x$, the model is: 
\begin{equation}
x_t = \beta_0 + \beta_1x_{t-1} + \beta_2x_{t-2} + \varepsilon_t^{(x)},
\end{equation}
where $\beta_0, \beta_1$ and $\beta_2$ are parameters that we estimate
using the standard least-squares method.
The benefit of a second-order model is its interpretability: it captures the physical principle that once the movement vector (direction and velocity) is determined, constant movement contains no information. 

After parameter fitting, we obtain residuals $r_t^{(x)}$ for each frame $t$:
\begin{equation}
r_t^{(x)} = x_t - \hat x_t = (\hat \beta_0 + \hat \beta_1\hat x_{t-1} + \hat \beta_2\hat x_{t-2})
\end{equation}
The residuals $r_t^{(x)}$ correspond to deviations from constant movement and are hence the part of the sequence unexplained by the autoregressive model. They can be used to gauge the surprisingness of the trajectory. The same procedure is carried out for sequence $y$.
We could now compute the differential entropy of a residual sequences, but as stated above, it alone has little meaning and we are in fact only interested in the mutual information between the two sequences.

Before we compute the mutual information, \emph{dimension reduction} is performed whereby multitouch gestures are represented using only as many features per measurement as the data requires.
The motivation for this step is that one cannot simply add information in the movement features in multitouch gestures together. Instead, any dependencies between the fingers should be removed.
Intuitively, a multitouch gesture with all fingers in a fixed constellation contains essentially the same amount of information as the same gesture performed using a single finger.
Dimension reduction is performed using principal component analysis (PCA), which removes any linear dependencies. %
Following common practice, the number of retained dimensions is set by finding the lowest number of dimensions that yields an acceptably low reprojection error (e.g., mean square error; see \cite{oulasvirta2013information}). 

Once the movement features have been processed by a dimension reduction technique (PCA), we treat them independently which amounts to simply adding up the information content in each feature in the end. Hence, the following discussion only considers the one-dimensional case where both $x$ and $y$ are univariate sequences.

Another issue in gesture data is that the two gestures $x$ and $y$ are often not of equal length due to different speed at which the
gestures are performed. This can be corrected by \emph{temporally aligning} the sequences using, for instance,
Canonical Time Warping \cite{zhou2009canonical}. The result is a pairwise alignment of each of the frames in $x$ and $y$ achieved
by duplicating some of the frames in each sequence. These duplicate frames are skipped when computing mutual information
to avoid an inflating their effect. 

Finally, we form pairs of residual values $(x_t,y_t)$ corresponding to each of the frames in the aligned residual sequences and evaluate the mutual information.
Since the mutual information is defined for a joint distribution of two random variables, we model the pairs $(x_t, y_t)$ of residuals in each frame $1 \leq t \leq n$ using a bivariate Gaussian model, under which the
mutual information is given by the simple formula
\begin{equation}
I(x;y) = -\frac{n}{2} \log_2(1-\rho_{x,y}^{2}),
\end{equation}
where $\rho_{x,y}$ is the Pearson correlation coefficient between $x$ and $y$.

By substituting the sample correlation coefficient estimated from the data in place of $\rho_{x,y}$ and subtracting a term due to the known statistical bias of the estimator (see \cite{oulasvirta2013information}), we obtain
the mutual information estimate
\begin{equation}
\hat I(x;y) = - \frac{n}{2} \log_2(1-r^{2}) - \log_2(e)/2,
\end{equation}
where $\log_2(e) \approx 1.443$ is the base-2 logarithm of the Euler constant.

The total information content in the gesture, based on two repetitions, is estimated as the sum of the mutual information estimates in each of the movement features after dimension reduction.

\textbf{Summary. } The metric has some promising properties to serve as an index of security. 
First, a distinctive feature of our framework that sets it apart from the work on symbolic password security is that in dealing with continuous gestures, it is imperative to be able to separate the complexity due to intended aspects of the gesture from that due to its
unintended, and hence non-reproducible, aspects. This is the
main motivation to use mutual information as a basis for the metric.
Second, as mutual information under the bivariate Gaussian model is determined by the correlation between the movement sequences (residuals), it is invariant under linear transformations such as change of scale, translation, or rotation. Hence, the user need not remember the exact scale, position, or orientation of the gesture on the screen. The metric is also independent of the size and the resolution of the used screen unless, of course, the resolution is so low that important details of the gesture are not recorded. 
Third, the time warping
step ensures that variation in the timing within the gesture has
only slight effect on the metric. %
Fourth, the metric enables comparison between gestures of unequal lengths and between single-finger and multi-finger gestures, as well as across different screen sizes and resolutions, on a unified scale (bits).

\section{Method}

Our study design builds on a generate-test-retest paradigm where participants were first asked to create a gesture, recall it, and recall again during the second session, a minimum of 10 days later. Participants were told that they should generate secure gesture as they would do in real situation and that their ability to recall them will be tested later. They were not given any hints about what a secure gesture might be. For understanding the generation and recall process, we used a mixed method approach: after generating gesture, all participants filled a questionnaire on workload (NASA-TLX \cite{hart:tlx}) after each task and a short survey in the end of second session.

We note that a somewhat similar generate-test-retest design has been used before by Chiasson et al.~\cite{Chiasson:2009:MPI:1653662.1653722} to compare multiple password inference to recall between text-based and graphical passwords (PassPoints \cite{Wiedenbeck:2005:PDL:1090412.1090418}). However, our work uses TLX forms, and is focused on free-form multitouch gestures, there are more repetitions and recalls, does not have a separate login phase, and the questions are asked at the end.

Next, we describe our volunteer participants, our apparatus, data preprocessing, experiment design and procedure.

\textbf{Participants.} We recruited participants with fliers, email lists, and in person in cafeterias. We required the participants to be 18 years old or over and familiar with touchscreen devices. %
We recruited 63 participants in all, from the ages of 18 to 65 (M = 27.2, SD = 9.9), 24 are male and 39 are female. Their educational background varies: 22 have high school diploma, 23 have a Bachelor's degree, 16 have a graduate degree and two have other degrees. %

All 63 participants completed session 1 of our study, and 57 of them returned and participated in session 2.
As compensation, participants received \$30 for completing the whole study.%
They also participated in a raffle of three \$75 gift cards.

We have recruited our participants in two batches: first in May 2013 (33) and second in June 2013 (30). %
Further, in order to analyze effect of varying time on recall, the gap between the two sessions of the study varies. The mean time gap for the first participants is 14.53 (SD = 5.81) days and 29.52 (SD = 7.57) days for the second.

Our study was approved by our Institutional Review Board.

\textbf{Apparatus.} The gesture data was recorded on a Google Nexus 10 tablet with Android 4.2.2 as the operating system, at an average of 200 frames per second (FPS).%

\textbf{Preprocessing.} The raw data files were preprocessed using MATLAB. %
In the preprocessing,  each gesture file is resampled to 60 FPS, to reduce the effect of the uneven sample rate.

The resampling was done by cubic spline interpolation, via MATLAB's built in function interp1d.m. For the x and y coordinates for each finger, it takes the recorded timestamps and resamples to a constant rate. The reduction from 200 to 60 FPS takes into account a large amount of duplicate data created by the touchscreen, and is necessary to prevent artifacts in the resampling. 

FFT analysis showed that the frequency reduction combined with the cubic spline method results in low pass filtering of the data, removing high frequency jitter introduced by the touchscreen hardware, but preserving the low frequency content of the gesture data. To deal with the uneven sample rate in the non-interpolated data, the Lomb-Scargle method was used. %

At this stage, artifacts in the raw data were detected and corrected as well. The primary such artifact is when a subject fails to place their fingers on the touchscreen in the same order between consecutive trials. As fingers are numbered sequentially upon being detected by the touchscreen, this places them out of order, resulting in artificially low scores in the later analysis. This was corrected by comparing the starting coordinates of each finger, and correcting the order to be consistent.

\textbf{Experiment Design.} The experiment followed a 17 x 2 mixed factor design with a repeated measurement variable of gesture repetition (17 levels) and a between-subject variable of time gap between sessions (2 levels). In the 17 gesture repetitions, 10 were performed during the creation process, followed  by another 2 repetitions after a short distraction, and 5 were performed in the second session.

\textbf{Procedure.} We conducted a two-session study. The second session was held after a minimum 10 days after the first session. The details of the procedure were as follows.

\textit{First session:} First, the participants were introduced to the study, which included reading and signing the consent form, discussion of their rights as participants and how they will be compensated. \textbf{Gesture Creation (Generate)}. Each participant was given the same tablet and was asked to create what they thought would be a secure gesture by drawing on it. The participants were asked to generate a gesture that they think others could not guess, but they could also remember later.The participants could retry until they felt satisfied with their gesture. Then the participant repeated the same gesture for additional nine times on the same tablet. Participants were presented with a blank screen for drawing their gestures. The application did not limit the number of fingers participants use to create the gesture. However, the number of fingers used could not be changed during the drawing process, that is, the gesture has to be drawn continuously without lifting any fingers from the screen.
Once it was completed, the gesture is displayed on the tablet's screen as a colored curve line, as shown in Figure~\ref{fig:example_photo}. The display was checked visually, to verify that the gesture was recorded properly. \textbf{Subjective Workload Assessment} The participant was asked to fill out NASA-TLX form regarding the creation process. \textbf{Distraction}. The participant was asked to perform a mental rotation task and count down from 20 to 0 in mind.
\textbf{Gesture Recall 1}. The participant was asked to recall the gesture by repeating it twice on the same tablet using the same application.
 \textbf{Demographic questions}. The participant was asked usual demographics questions that were aggregated and reported above.

\textit{Second session:}  \textbf{Gesture Recall 2}. The participant was asked to recall his/her gesture by repeating it for five times on the same tablet using the same application.  \textbf{Subjective Workload Assessment.} The participant was asked to fill out NASA-TLX form regarding the recall process. \textbf{Short Survey}. The participants were asked some questions about the recall process and other thoughts about the study.

\newpage

\section{Results}

For each participant who completed both sessions, a total of 17 gesture repetitions were recorded. In all, 1038 recordings were generated, as 6 participants did not attend session 2, and 3 additional traces were not completed. The groups of repetitions are summarized in Table \ref{tab:RepetitionGroups}. Because the estimated Mutual Information ($\hat I$), is computed in pairs, $\hat I$ is reported as the mean for the relevant repetitions.

\vspace{-5 pt}
\begin{table}[!h]
\begin{center}
\begin{tabular}{ccc}
    Session & Trial \# & Group \\\hline 
     1 & 1-10 & Generate \\
     1 & 11-12 & Recall1 \\
     2 & 13-17 & Recall2 \\
\end{tabular}

\end{center}
\vspace{-15 pt}
\caption{Repetition Groups by Session \# and Trial \#. $\hat I$ was computed for all pairs of repetitions each group.}
\label{tab:RepetitionGroups}
\end{table}
\vspace{-7 pt}

\subsection{Factors Affecting Security}

 \begin{figure}[!h]
  \centering
    \includegraphics[scale=0.5]{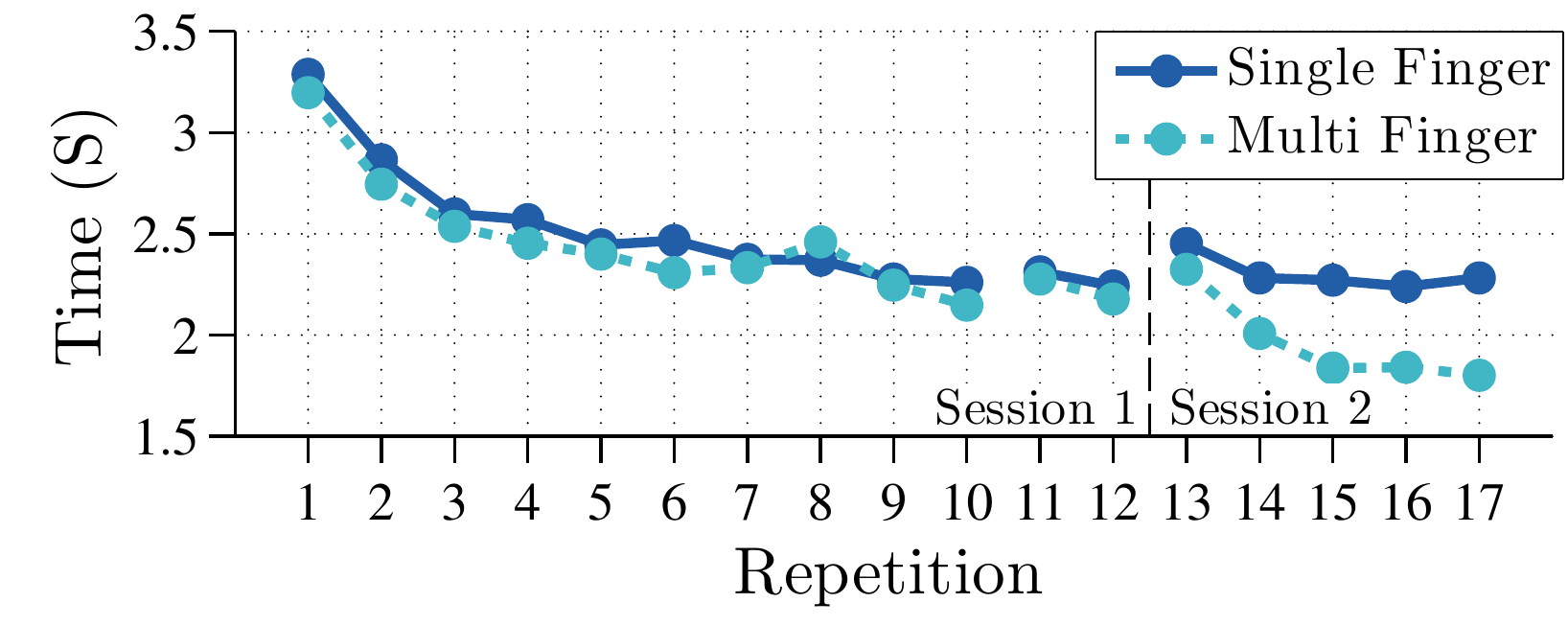}
    \includegraphics[scale=0.5]{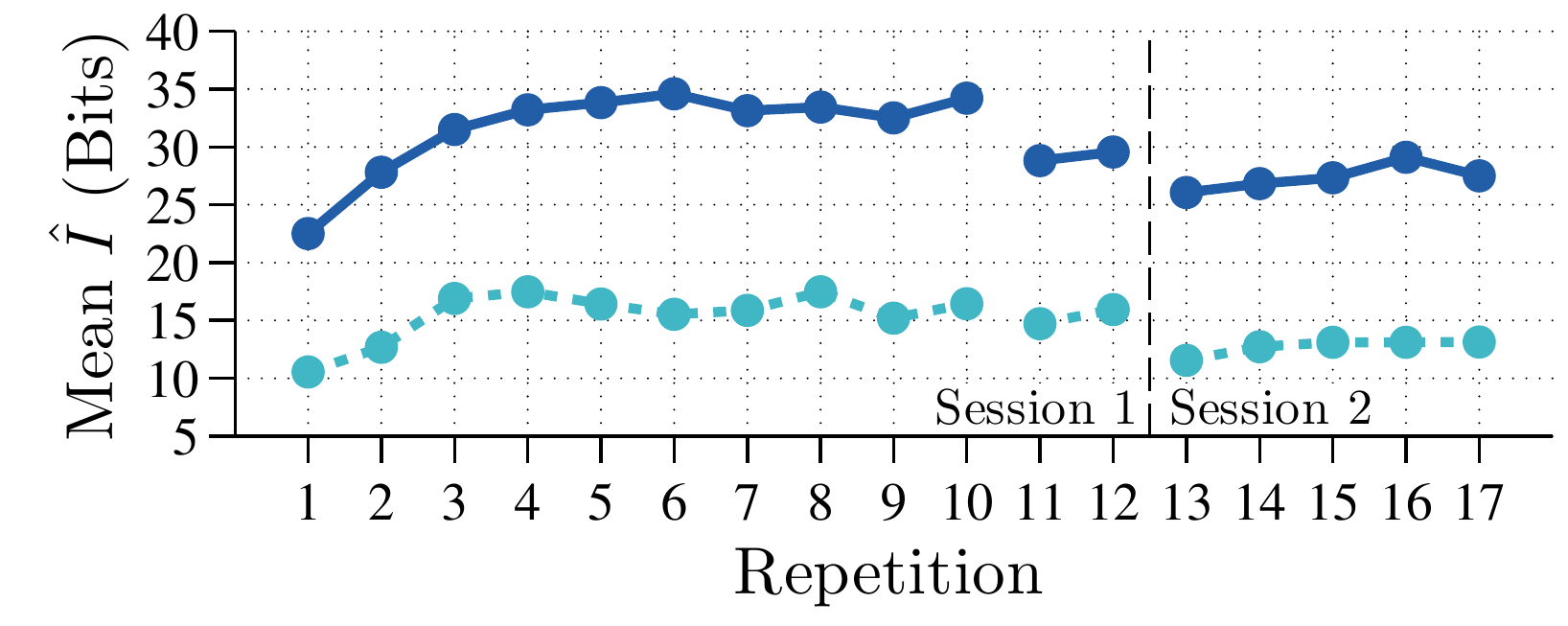}

  \caption{Mean  $\hat{I}$ and mean gesture duration vs Repetition. Top: Within Generate, mean gesture duration trended downwards as gestures increased in speed. Bottom: Over the same repetitions,  $\hat{I}$ trended upwards before leveling out. It then dropped quickly between Generate and the two Recalls. }
  \label{fig:TimevsRep1}
  \label{fig:SIvsRep1}

  \end{figure}

Figure \ref{fig:TimevsRep1} shows the mean $\hat{I}$ of each repetition across all gestures versus the repetition number. During gesture creation in Generate, $\hat{I}$ trended upwards from repetitions 1-4, and then leveled off from repetitions 5-10. This shows that it takes at least three repetitions for a participant's gesture to become stable.
  
A second major feature is that by Recall1, repetition 11 and 12, the $\hat{I}$ has dropped suddenly, despite a delay of only a few minutes. Surprisingly, more than 10 days later, $\hat{I}$ did not drop much further in Recall2. The drop between Generate and Recall1 was more severe for single finger gestures, and the drop between Recalls 1 and 2 was more severe for multifinger gestures. In both cases, $\hat{I}$ stabilized, at around 27 bits for single finger and 15 bits for multifinger gestures, having dropped from an initial value of around 35 and 20 bits respectively.

Figure \ref{fig:TimevsRep1} also shows the amount of time taken to record each repetition of each gesture. This duration also changed with repetition. As the number of repetitions increased, the mean duration of each repetition trended downwards from around 3 seconds to around 2. Unlike $\hat{I}$, it remained stable from there through the two recalls. Interestingly, during Recall2 the multifinger gestures sped up from 2.5 seconds to under 2 seconds, whereas the single finger gestures stabilized.

\begin{figure}[!h]
  \centering
  \includegraphics[scale=0.5]{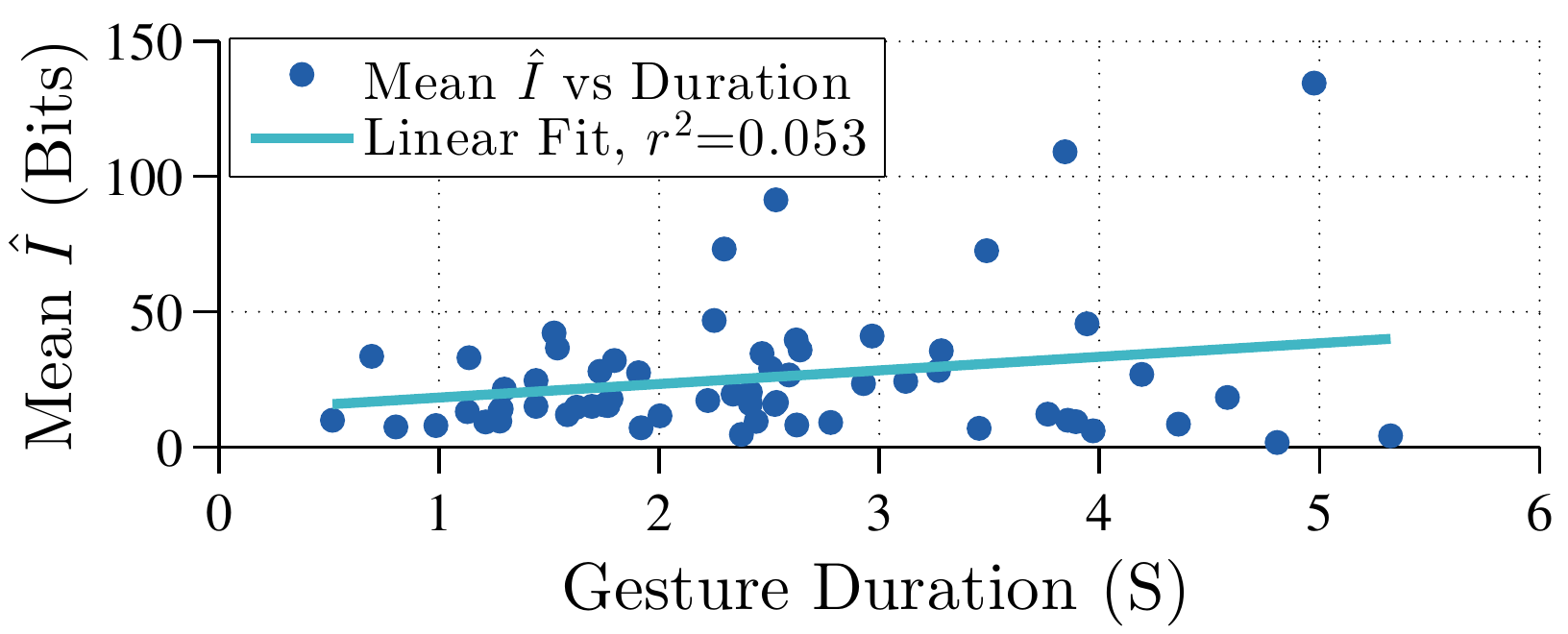}
 \caption{Mean  $\hat{I}$  vs. mean gesture duration. The r\textsuperscript{2} value indicates a poor linear fit, showing little correlation. However, the highest $\hat{I}$ gestures were all longer than 2 seconds, suggesting that some degree of precision is required.}
  \label{fig:SIvsTime1}
  \end{figure}

A plot of the mean  $\hat{I}$ of each gesture versus mean duration appears in Figure \ref{fig:SIvsTime1}. This shows that many gestures with a short duration also had a low  $\hat{I}$. The highest  $\hat{I}$ gestures had a duration of between 2 and 5 seconds.  $\hat{I}$ increased with duration, but had a poor fit, explaining only 5\% of the variation, as the highest duration gestures were either very high or very low  $\hat{I}$. Long duration could thus indicate either a complex, careful gesture, or a relative lack of practice.

\begin{figure}[!h]
  \centering
  \includegraphics[scale=0.5]{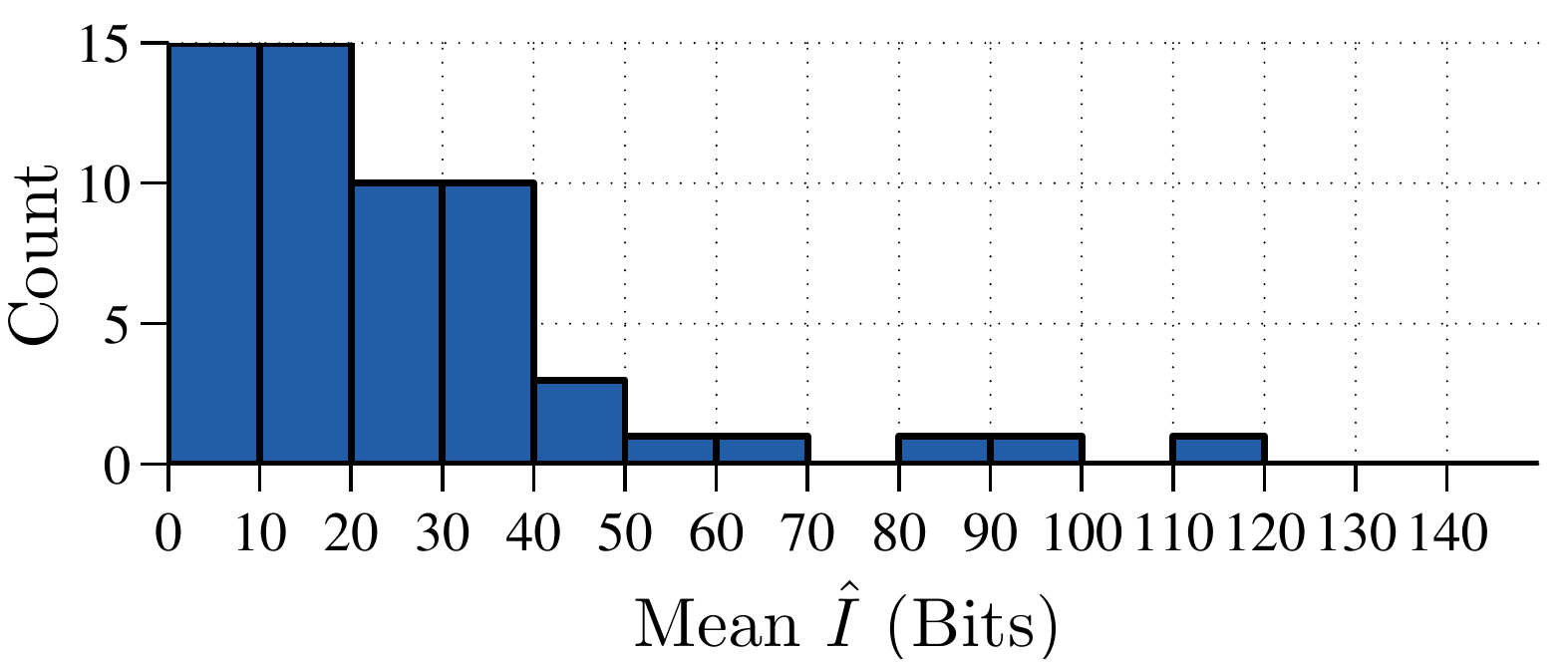}
    \caption{Histogram of mean  $\hat{I}$ of Generate, per gesture, showing low scoring, biased distribution.}
  \label{fig:histSI}
  \end{figure}

Figure \ref{fig:histSI} shows that majority of the user-generated gestures had relatively low $\hat{I}$ with a only a small tail of high $\hat{I}$. For Generate, the distribution had a mean of 27.72 bits, and a standard deviation of 26.30 bits. However, taking only the second, stable half of Generate, the mean rose to 33.42 bits, with a standard deviation of 31.13 bits. In both cases, the standard deviation was about the same size as the mean, indicating a large variability. The histogram shifted as well, becoming slightly more uniform. Although many user chosen gestures score poorly, some scored highly as well, suggesting that it is possible create guidelines to emulate the characteristics of high scoring gestures.

We compared mean $\hat{I}$ with participant age and gender. There was a mild negative relationship between $\hat{I}$ and age, with a r\textsuperscript{2} of 0.08. Seven of the top eight gestures were created by participants under the age of 25, with the remaining one under the age of 30. %
Mean $\hat{I}$ for male participants (N=23) was 29.82 bits and mean $\hat{I}$ for female participants (N=40) was 25.00 bits.

\begin{figure}[!h]
\begin{center}
\includegraphics[scale=0.5]{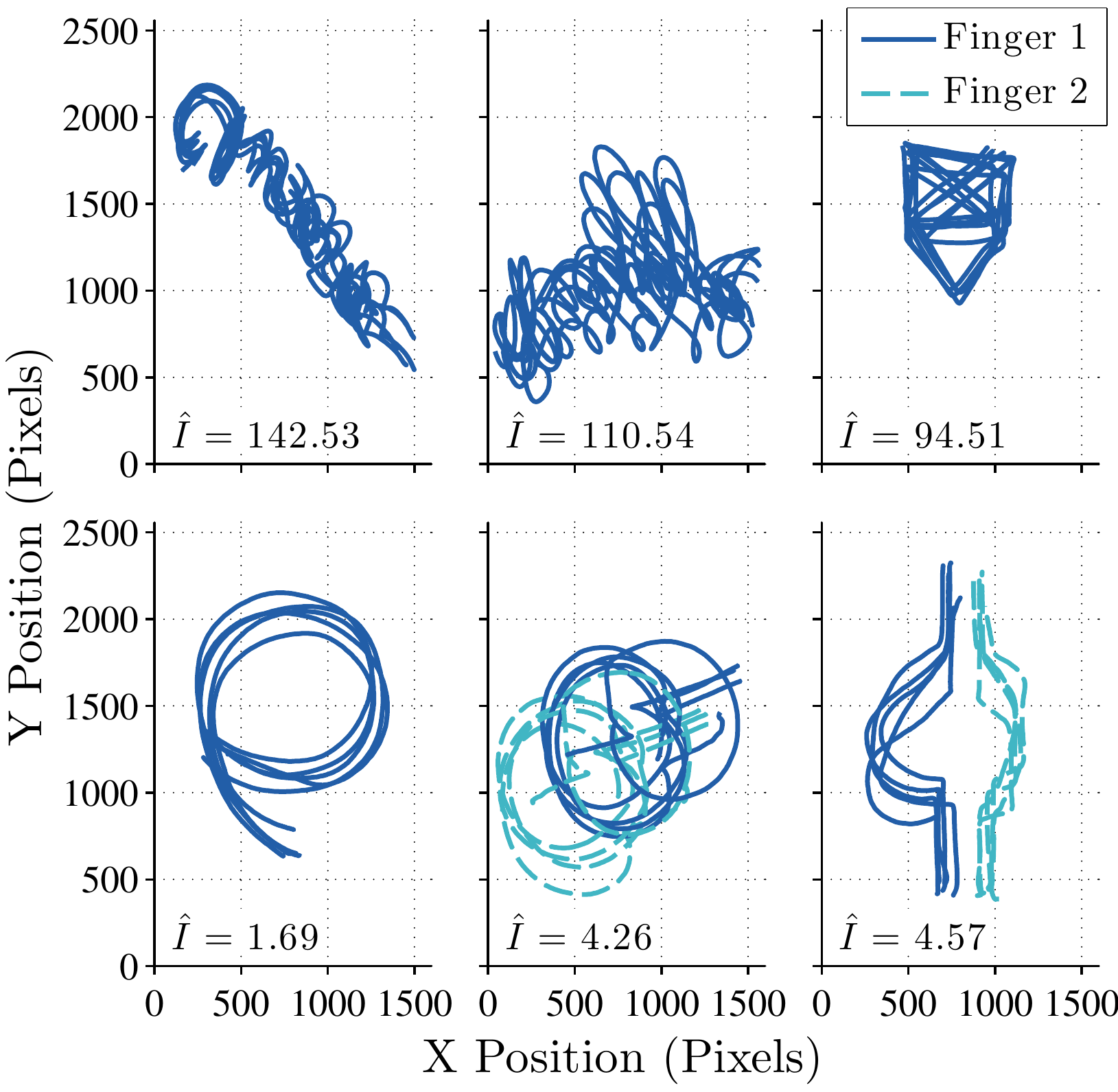}
\caption{Gestures ranked by $\hat{I}$ for Generate and Recall2. Top: Best three gestures, showing the many tight turns characteristic of high scoring gestures. Bottom: Worst three gestures, Low scoring gestures had few, gentle turns.}
\label{fig:gallerysummary}
\end{center}
\end{figure}

Finally, we looked for defining visual characteristics of the highest and lowest scoring gestures. We ranked each gesture by its $\hat{I}$ in each of the five categories, and evaluated the top five in each category. There was high correlation between the categories, and as such, the top five gestures for each category overlapped significantly, having only nine unique gestures. The best gestures fell into two groups, angular paths with many hard turns, and signatures. This matched our expectations, as the algorithm looks for both consistency between trials and for large deviations from a straight line. The defining visual feature of the lowest scoring gestures was having only a few, gentle curves. Many were multifinger, with the additional fingers merely copying the motion of the first.  A gallery  appears in Figure \ref{fig:gallerysummary}.

\newpage
\subsection{Security of Multitouch Gestures}

\begin{figure}[!h]
  \centering
  \includegraphics[scale=0.5]{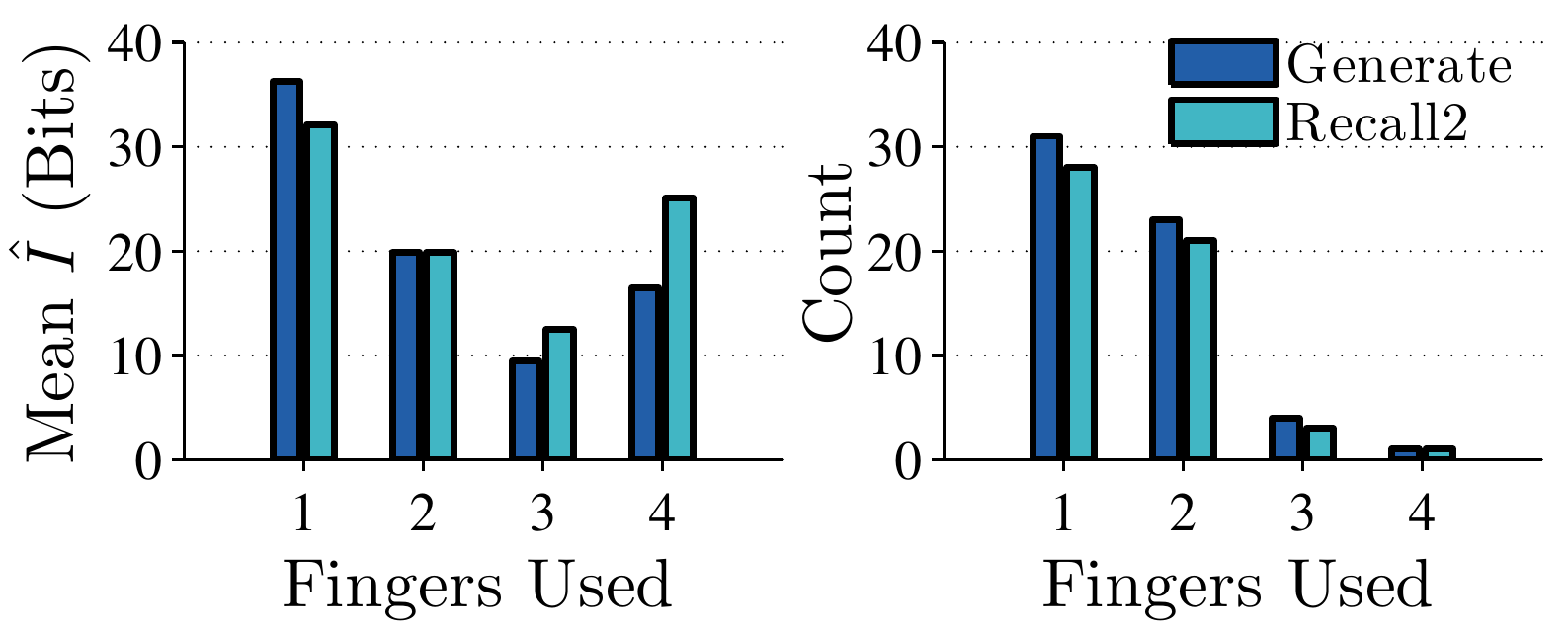}
  \caption{ $\hat{I}$ and number of fingers used to perform gesture, and change between sessions 1 and 2. This shows the both the higher performance of single finger gestures, as well as the rarity of using more than two fingers.}
  \label{fig:barFingers}
  \end{figure}

  As seen in Figure \ref{fig:SIvsRep1}, the mean $\hat{I}$ of multifinger gestures is  lower than that of single finger ones. We compared $\hat{I}$ of these gestures to estimate how much additional information is added by additional fingers. Figure~\ref{fig:barFingers} shows the higher mean $\hat{I}$ of single finger gestures, and the rarity of gestures using more than two fingers. Recall2 showed a greater difference in  $\hat{I}$ than Generate, as a number of participants failed to use the same number of fingers when they returned in session 2. Of the 63 participants, 32 decided to create multifinger gestures, and 31 chose to create single finger gestures, with only three participants using more than two fingers. Participants were prompted that they could use as many fingers as they liked, but were not instructed on how many to use.

 We performed regression analysis on the effect of the number of fingers  on  $\hat{I}$. The result shows that the effect is significant for the  $\hat{I}$ of Generate, $b=17.948, t(57)=2.763, p=.0077$, while not for the  $\hat{I}$ of Recall2, $b=11.898, t(57)=1.841, p=.07$. However, the regression model only explained 11.8\% of variance in the  $\hat{I}$ for the significant effect. In short, the number of fingers is not the most major factor effecting $\hat{I}$.

\subsection{Factors Affecting Memorability}

\begin{figure}[!h]
\begin{center}
\includegraphics[scale=0.5]{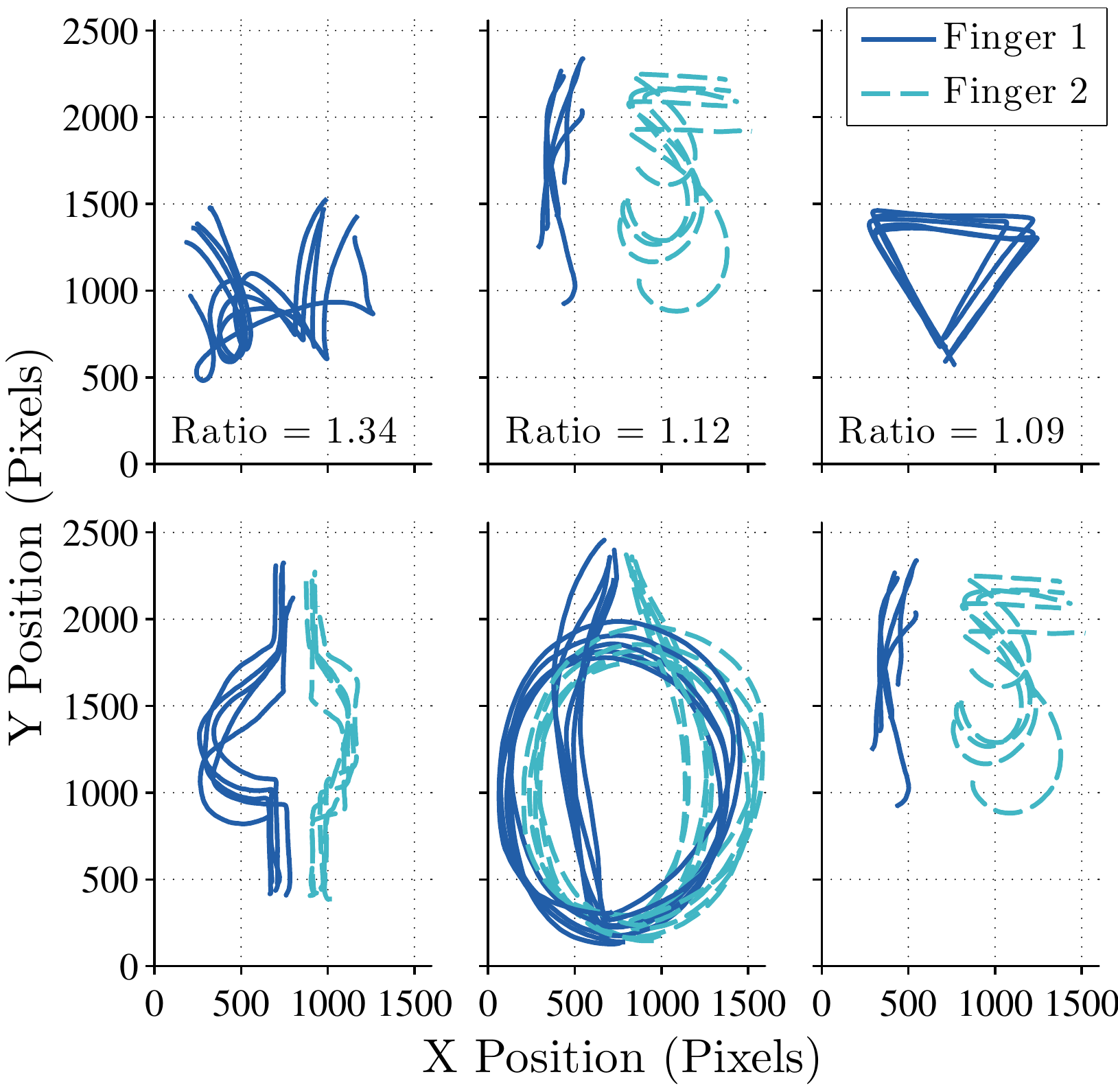}
\caption{Top: Best 3 Gestures by memorability. These have a shorter length and decreased complexity compared to high  $\hat{I}$ gestures. Bottom: These 3 gestures had the greatest difference in path between fingers. Two of the three are a simple mirroring of the path (Left, Middle), while only (Right) adds a large amount of  $\hat{I}$.}
\label{fig:MemoryRatio}
\label{fig:FDiff1}
\end{center}
\end{figure}

Figure \ref{fig:MemoryRatio} shows the best remembered gestures. To evaluate memorability, we computed the cross group $\hat{I}$ of Generate and Recall2, with pairs consisting of one from each group, instead of both from within a group. However, the large differences in $\hat{I}$ from gesture complexity obscured the differences from repetition accuracy, as the cross group $\hat{I}$ has a linear fit with an r\textsuperscript{2} of 0.65 with the $\hat{I}$ of Generate. We compensated by dividing the mean cross group $\hat{I}$ for each gesture by its $\hat{I}$ from Generate. 

Gestures that scored highly on the resulting ratio have the best consistency between Generate and Recall2, as compared to the consistency within Generate. The top gestures for memorability are shorter and simpler than the top gestures for security.

\begin{figure}[!h]
\centering
\includegraphics[scale=0.5]{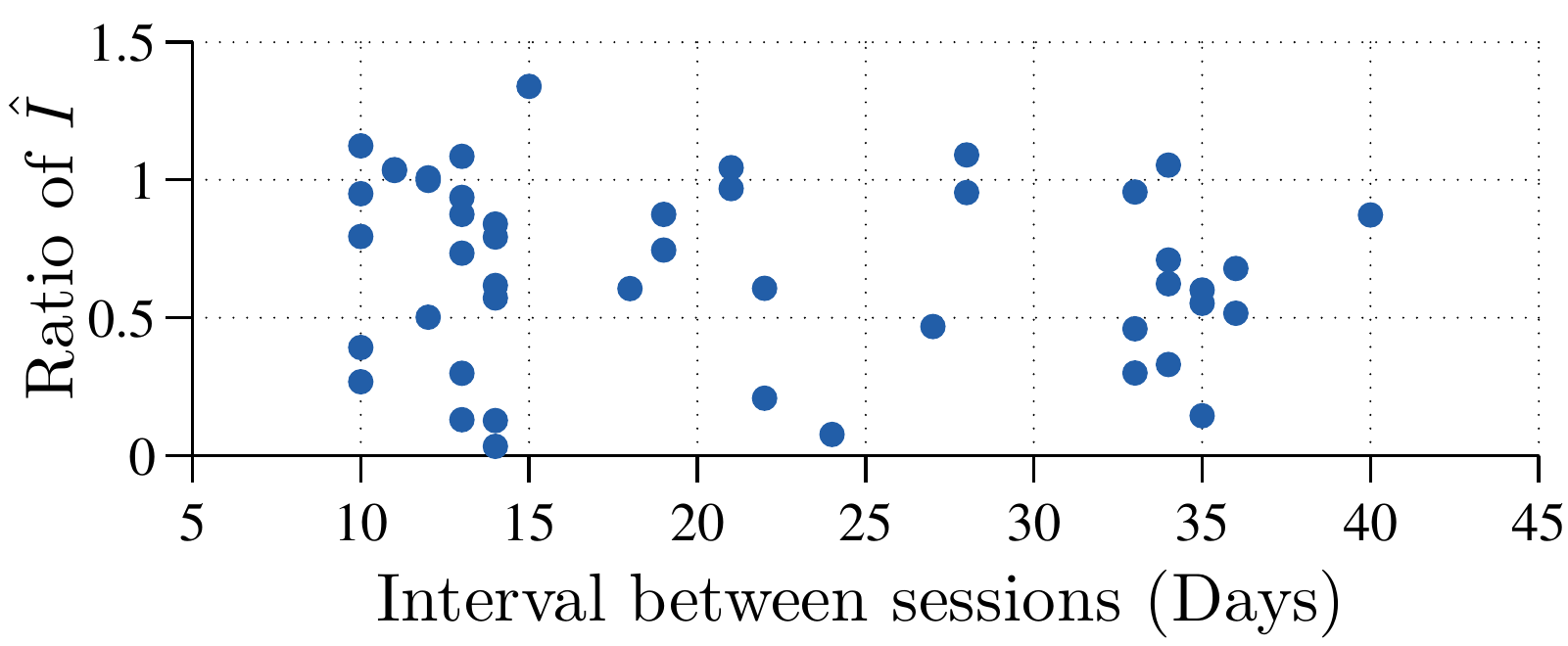}
\caption{Memorability vs interval between sessions. This plot compares the relative quality of gesture recall versus time between Generate and Recall2. There was no significant correlation, with a  (r\textsuperscript{2} < 0.03).}
\label{fig:delaySI1}
\end{figure}

Once we had a way of comparing how well gestures are remembered, we investigated what might cause the large difference in $\hat{I}$ between sessions. We compared the memorability ratio to the time interval between Generate and Recall2, as seen in Figure \ref{fig:delaySI1}. The linear fit however, had a r\textsuperscript{2} of less than 0.03, showing minimal dependence on the delay between sessions.

\subsection{Individual Differences}

  Given the surprising deficiency of the multifinger gestures, we looked at specific examples. Only three participants used multitouch gestures with significantly different motions between fingers, and in two of the three cases they were simple mirrorings of the motion. These three gestures appear in Figure \ref{fig:FDiff1}. All other cases were just translations of the same trace, as if the gesture were made with a rigid hand. Gestures with minimal additional information per finger were in part scored low because all gestures were run through a PCA algorithm to remove redundant information, prior to analysis of $\hat{I}$.

Participants also commonly performed several categories of error. Despite instructions, 19 of the 32 multifinger using participants placed their fingers down in an inconsistent order between sequential trials. This was fixed in preprocessing to ensure that computation of $\hat{I}$ used the same fingers. In addition, some participants rotated the tablet either during a session or between sessions. This stands out visually when comparing traces, but the algorithm is negligibly affected by starting angle, as it is looking at the residual as the trace turns from its origin. This was verified by comparing the result with its session one or two counterpart that did not have said rotated trace. Some of the lowest scoring gestures featured a rotation between repetitions, and all gestures with rotation between sessions scored poorly on cross session $\hat{I}$. However, the gestures with cross session rotation also scored poorly within a session. Only one gesture with rotation between sessions scored significantly higher on its within session $\hat{I}$ than on its cross session.

Given that error containing gestures scored poorly, even when excluding those repetition pairs containing the errors, we take these errors as an indicator of poor recall. Extra fingers or complex shapes are no guarantees of a high score, without consistent execution. Attention when creating the gesture is thus important, practicing accurate repetition rather than just going through the motions.

\subsection{Subjective Task Load Assessment \newline Analysis}

\begin{figure}[!t]
\begin{center}

\includegraphics[width=0.9\linewidth]{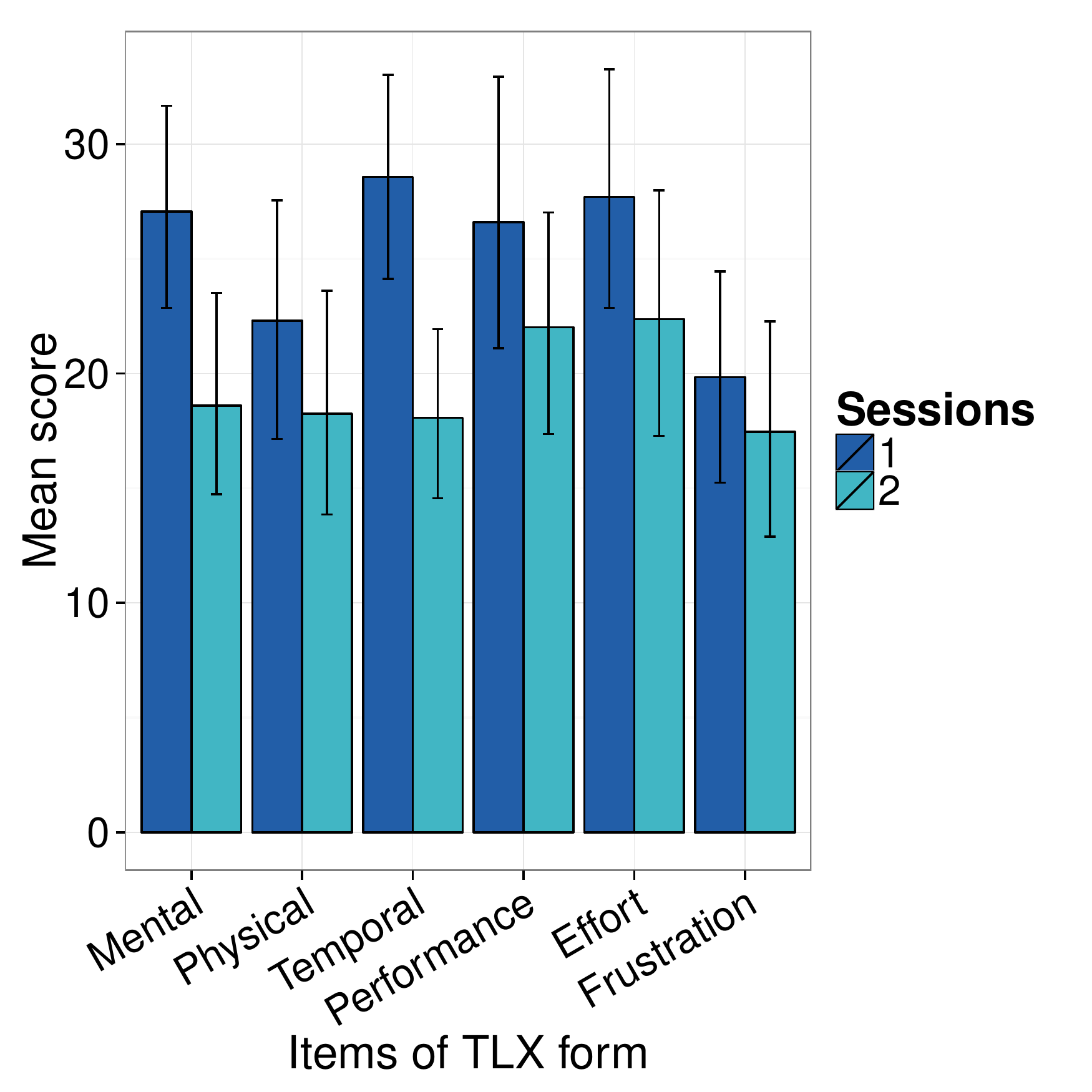}
\caption{\label{fig:tlx_by_item_by_session} Mean score and corresponding 95\% confidence interval for each item on the TLX form, for the two sessions of our study. The mean scores of each item in session two were lower than those of session one.}
\vspace{-15pt}
\end{center}
\end{figure}

This section contains the analysis of the study's TLX forms. For each session of our study we asked participants to fill out one TLX form, which is used to assess subjective workload of given task. Figure~\ref{fig:tlx_by_item_by_session} shows the mean score and corresponding 95\% confidence interval for each item in the TLX form, for both sessions. 

\begin{table}[!h]
\begin{center}
  \begin{tabular}{ l | c  c  c}
  \hline
	Item & Mental & Physical & Temporal\\ \hline
	$p$ value & $<$0.001 & 0.025 & $<$0.001\\
	effect size & -.038 & -0.21 & -0.34\\	\hline
	Item & Performance & Effort & Frustration\\ \hline
	$p$ value & 0.029 & 0.0013 & 0.072 \\
	effect size & -0.20 & -0.30 & -0.17 \\	
  \end{tabular}
  \caption{Wilcoxon signed-rank test on the scores of each item of the TLX result for the two sessions, showing significant differences in all except Frustration.}
    \label{tab:wilcoxon_tlx}
    \vspace{-11pt}
\end{center}
\end{table}

Figure~\ref{fig:tlx_by_item_by_session} suggests that the mean scores of session two are lower than those of session one. To prove the point we conducted a non-parametric repeated-measure Wilcoxon signed-rank test on the scores of each item from the two sessions, given the fact that the data does not follow a normal distribution. The result is shown in Table~\ref{tab:wilcoxon_tlx}.

From Table~\ref{tab:wilcoxon_tlx} we can see that except for Frustration, all items show significant differences in scores between sessions. It is safe to say that given the data we have, it is very likely that participants felt the recall task was easier than the creation task in terms of workload. This could be because of increased practice or familiarity, and intuitively agrees with the increase in gesture speed seen in Figure \ref{fig:TimevsRep1}.

\section{Practical Authentication \\ System Implementation}

In this section, we describe our extension to a practical
single touch gesture recognizer for multitouch gestures, and
the results of how the participants' gestures would perform in
a real authentication system based on our multitouch recognizer.
We also present our trial on shoulder surfing attacks that indicates
how free-form gestures are robust against shoulder surfing.

A recognizer works by taking a user's gesture,  passing it through a recognition algorithm, and computing whether or not the gesture is a successful match for a stored template. The device will store a series of templates of which the gestures are compared to for authentication; the best score is used and compared to a threshold value. We have the following assumptions for the recognizer: 1) location invariance: No matter where the correct gesture is drawn on the screen, it should be authenticated correctly.  2) scale invariance: No matter what size the correct gesture is drawn to on the screen, it should be authenticated correctly. 3) rotation invariance: No matter what angle the correct gesture is drawn at on the screen, it should be authenticated correctly. Location and scale invariance are important when dealing with cross-platform authentication; the screen dimension inherently limits what size the gesture can be drawn to and the area over which a gesture can be performed would cause wild variations in where it would be drawn depending on the user. Rotation invariance is useful for reducing computational complexity when dealing with individualized free-form gestures as we have in our data set. We note that authentication system designers can opt to restrict or relax these assumptions.

	We elected to implement and extend the Protractor \cite{Li:2010:PFA:1753326.1753654} recognition algorithm, a popular nearest neighbor approach. Given the gesture templates obtained and the two recall sets, we would like to measure how well the gestures perform. Protractor is an improvement upon the \$1 Recognizer \cite{Wobbrock:2007:GWL:1294211.1294238}, having both a lower error rate \cite{Li:2010:PFA:1753326.1753654} and an effectively constant computational time per training sample as compared to \$1's growing cost per training sample. Protractor presents itself further as an attractive algorithm for the data under consideration since it has low computational complexity compared to other techniques, for example, Dynamic Time Warping (DTW) \cite{Wobbrock:2007:GWL:1294211.1294238} and Hidden Markov Models (HMM) \cite{hmm1,hmm2}.  In general, Protractor's error rate falls with an increasing number of training samples and at 9-10, the error rate is less than 0.5\% \cite{Li:2010:PFA:1753326.1753654}. 

	Protractor is only a single touch recognition algorithm; other projects considering gestures have used more general techniques for gesture recognition instead of a nearest neighbor approach. For example, Sae-bae et al. \cite{Sae-Bae:2012:BGN:2207676.2208543} applied DTW to deal with their multitouch gesture set. For the authenticator to remain practical, it needs low computational complexity, high speed,  and low error rate per template to be implemented on a mobile device -- Protractor can meet this demand; DTW cannot. As we are dealing with multitouch gestures, it is necessary to modify Protractor. The accounting procedure is as follows.

\begin{enumerate}
\item[] \textbf{Our Multitouch Extension to Protractor}
\item  Each finger is split into its own set of points and passed through the algorithm and compared to templates of similar fingers and the score is computed for each individually. 
\item They are then averaged together.
\item There are provisions built into place to ensure the authentication failure for the wrong number of fingers. In the case of n fingers versus m, the number of fingers is compared. If n is equal to m, then the recognizer continues to the next step. If n is not equal to m, then the recognizer immediately stops the computation and registers the score as 0; a failure.
\item This score is then  is compared to the threshold value. If the score is greater than or equal to the threshold then it is considered a positive authentication; otherwise, it is negative.
\end{enumerate}
It is important to note that the threshold should be set high enough such that authentication failure is all but guaranteed for gestures that are being matched to templates other than their own.

       As a reminder, when the participants began the study, they were asked to repeat their gestures ten times; each of these ten trials is used by Protractor as templates for that gesture. %
There are two authentication data sets under consideration here: the first is where participants were asked to replicate their gesture after a mental task and the second where they were asked to replicate their gesture after at least 10 days.

\subsection{Recognizer Performance}

\begin{figure}[t!]

\includegraphics[scale=0.4]{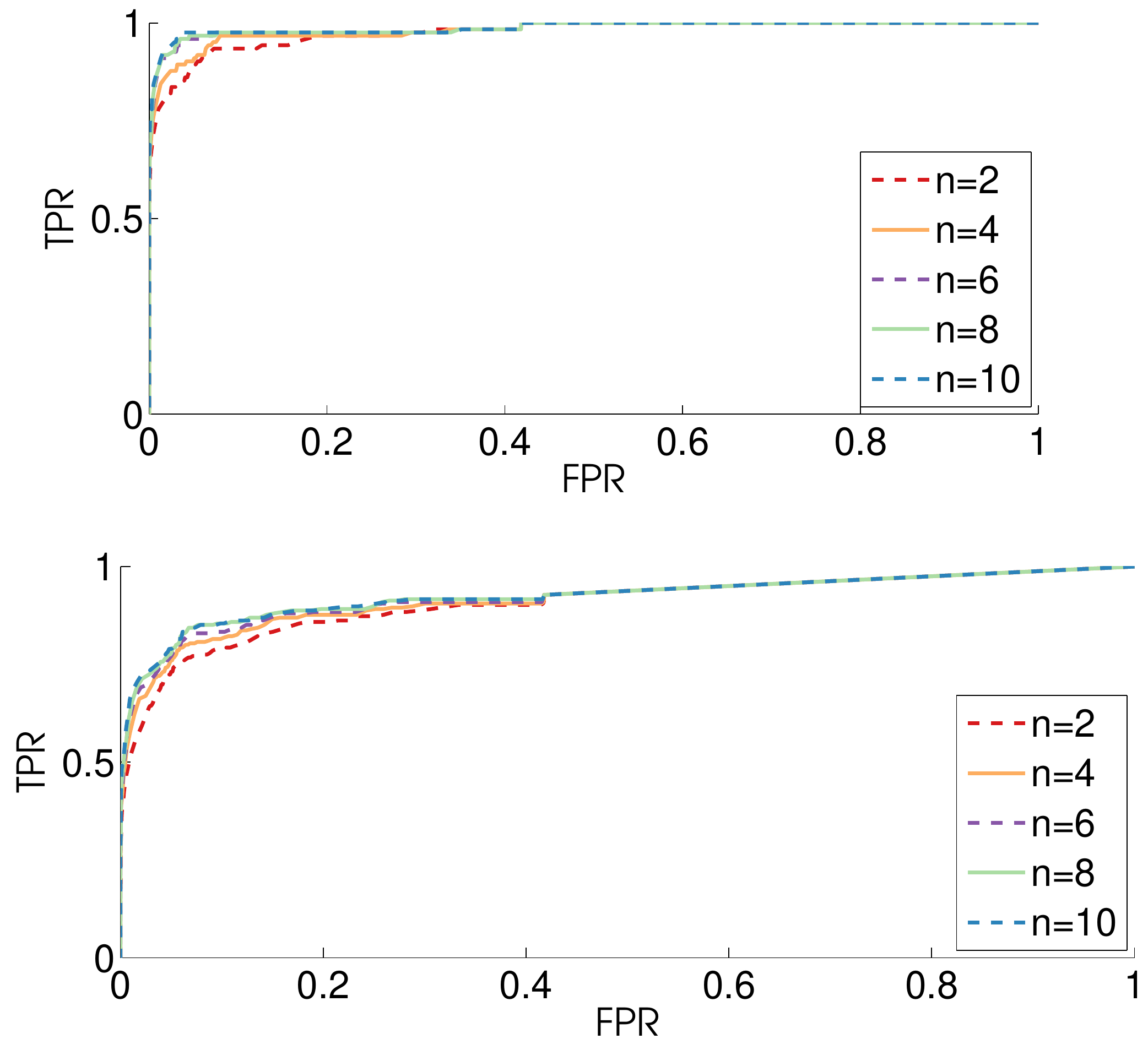}
\caption{This figure shows the ROC curve for the recognizer across the two data sets with variable template numbers --   'n' corresponds to the number of templates. The top plot is the first data set and the second plot is the second data set. The ROC curve is a measure of the performance of a binary classifier; the closer the top left corner of the plot moves towards the vertical axis, the better the performance. The first data set is closer to that corner than the second, showing that the second data set performed worse and must have a higher equivalent error rate. As the number of templates increases, the closer the curve shifts up and the better the recognizer performs. In general, the recognizer classified the first data set with a lower EER than the second set despite those being the same gesture types, indicating a weakness on the parts of participants to accurately replicate their gestures. The EER values can be read in Table~\ref{tab:eerges}.}
\label{fig:roc1}
\end{figure}

We wanted to know how accurately the recognizer is performing across the different gestures in our data sets. To quantify this in terms of a numerical estimate and visualize it, we elected to obtain a Receiver Operating Characteristic (ROC) curve and derive from it an Equivalent Error Rate (EER). The ROC curve gives the visual representation of how our classifier is performing and the EER value gives us the numerical estimate for how it is performing; the lower the EER is, the more accurately the recognizer is performing.

To find the EER value we need to find the rate at which the True Positive Rate (TPR) is equivalent to the False Positive Rate (FPR). These are defined as:

\begin{equation}
\label{tpr}
TPR = \frac{True\:Positives}{True\:Positives + False\:Negatives}
\end{equation}

\begin{equation}
\label{fpr}
FPR = \frac{False\:Positives}{False\:Positives + True\:Negatives}
\end{equation}

These values are dependent on the threshold parameter that the score computed by the recognizer is compared to. To generate an ROC curve, one must vary the authentication threshold parameter and measure a (TPR,FPR) value for each point and plot them. From there, we can determine the EER visually. 

At low threshold values, the classification system would be accepting virtually any gesture as a validation against any template gesture. At high values, all input gestures would be rejected against any other template (even if it is being matched to the correct one). %
 As the EER value climbs higher, the reliability of the gesture or the recognizer can be called into question since false positives are propagating through the system.  %
 In the case of our system, the templates are not matched to one another. Instead, we have two true cases for a given gesture in the first data set and five true cases for a given gesture in the second dataset. All other gesture attempts in all other datasets (exclusively across these two sets; no intersections) are considered false cases. 

\begin{table}[h]
\begin{center}
\begin{tabular}{c|cc}
\# of Templates & Set 1 EER& Set 2 EER \\
\hline
 2 & 7.07\% & 15.97\% \\
4 & 6.42\% & 14.45\% \\
6 & 4.13\% & 13.94\% \\
8 & 4.10\% & 13.09\% \\
10 & 3.34\% & 13.16\% \\

\end{tabular}
\end{center}
\caption{EER Values, Ranked by Template Number. Listed above are the EER values corresponding to Figure~\ref{fig:roc1}. As the number of templates increases, the lower the EER value drops and thus the lower the error in the system and the better the recognizer performs. The lower Set 2 values correlate to the shape of the curves represented in Figure~\ref{fig:roc1}; as EER decreases, the better the curve appears. The EER values for the recognizer reduce more slowly with 6 training templates, indicating this to be the ideal starting point when asking a user to train the system.}

\label{tab:eerges}
\end{table}

As for how well the gestures and the recognizer performed in terms of accuracy across the data sets, that information can be gleaned from  the ROC plots given in Figure~\ref{fig:roc1} and the EER values shown in Table~\ref{tab:eerges}. As a reminder, the further away an ROC curve moves from being a 90-degree box (an EER of 0\%), the worse it is performing. Figure~\ref{fig:roc1} shows the ROC curves with a varying number of template sizes. As the number of templates increases, the ROC curves are pushed further towards the corner and the EER values are lowered, telling us that a larger number of training templates leads to improved accuracy. For the first set, the best result (with 10 templates) is 3.34\% and the best result for the second set is 13.16\%. Note that the higher EER on the second data set does not speak to the weakness of the recognizer but rather those of the participants -- some participants who returned to attempt their gestures in the second set forgot the number of fingers they used, registering immediate authentication failure. As such, the increased error in the second set as compared from the first set can be attributed to recall problems rather than weaknesses in the recognizer's ability to classify gestures.

\subsection{Shoulder Surfing Attack Trial}

We conducted a preliminary study to understand how free-form gestures would resist shoulder surfing attacks.
Towards this end, we recruited seven participants from computer science and engineering schools who had considerable experience with touchscreens. We assume that these volunteers would likely to be more skilled with attacks than the general populace to limit confounding factors. One of the seven volunteers acted as the target who performs gestures and the other six would be attackers who try to replicate target's gestures.

\begin{figure}[tbph]
\begin{center}

\includegraphics[scale=0.5]{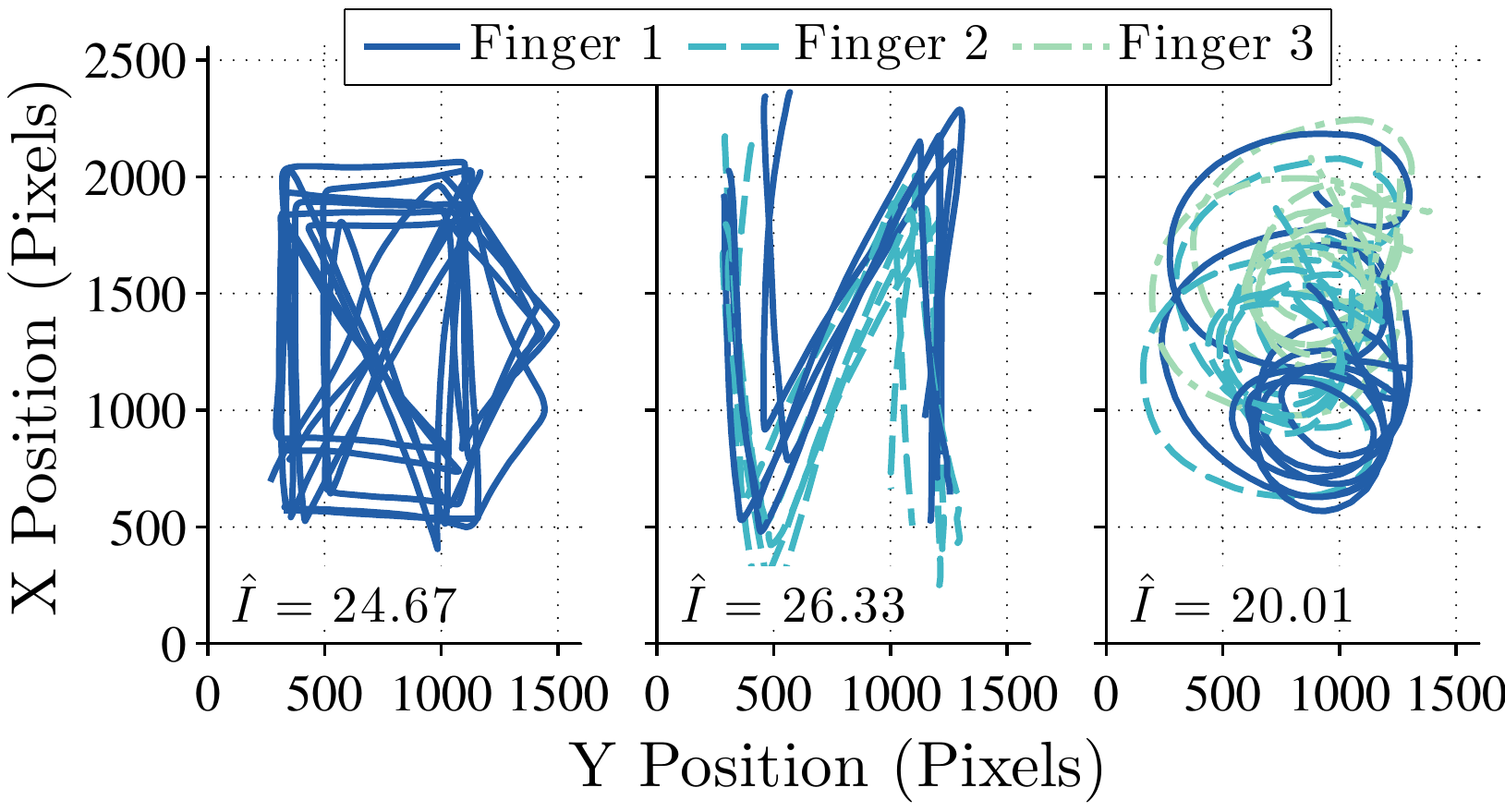}
\caption{Gallery of Gestures used for shoulder surfing. In order from left to right represents Gesture1, Gesture2, and Gesture3 generated by the Target.}
\label{fig:ShoulderSurfGallery}
\end{center}
\end{figure}

We chose three qualitatively different gestures as shown in Figure~\ref{fig:ShoulderSurfGallery} as examples.
In the experiment, we first had the target of the attacks exercise and get familiar with all three gestures; then we collected gesture data from the target in a way matching the original dataset: for each chosen gesture, the target first repeated 10 times; after a short distraction task, which included mental rotation and countdown, the target repeated another two times. Finally, we video recorded one additional repetition the target made for each gesture. Instead of having the target performing gestures in person for every attacker, we played video recordings of that process to the attackers, which ensured attackers would not be affected by any inconsistency or difference within the performance of the target.

During the shoulder surfing process, each attacker was presented with all of the three videos, each of which contained one of chosen gestures. The attackers were seated always at the same spot, adjacent to the a chair at the table where the display was setup. This was done in order to emulate shoulder surfing. The order of videos played to each attacker was produced in a Latin square to prevent any undesired effect on the performance of attackers. Each video was played only once for each attacker. Then attackers were told to repeat the gesture they observed from the video for five times with the purpose of replicating it as well as possible.

We measured shoulder surfing effectiveness of a gesture using our multitouch recognizer discussed above. The templates we used are the 10 gestures from the  Generate phase of the target. Table~\ref{tab:ShoulderSurfingReg} shows the result. All scores displayed in the table are maximum score of that category, i.e. the best attempt of either the target or attackers to replicate the gesture. Score that are rendered as a 0 are because of the fact that the recognizer immediately rejects template-gesture matching where the finger count is different. 

\begin{table}[h]
\begin{center}
\begin{tabular}{l|ccc}
Participant & Gesture1 & Gesture2 &Gesture3 \\\hline
Target (Recall) 	 & 4.36	 & 4.31 	 & 6.75 \\ 
Target (Video) 	 & 2.95 	 & 4.51 	 & 7.27 \\ 
Attacker 1 	 & 0.50	  &  0.00 	 &  0.97 \\ 
Attacker 2 	 & 0.43 	 &  3.08	 &  0.00 \\ 
Attacker 3 	 & 1.07 	 &  0.00 	 &  0.00 \\ 
Attacker 4 	 & 0.27 	 &  0.00 	 &  0.94 \\ 
Attacker 5 	 & 1.10 	 &  2.19 	 &  3.96 \\ 
Attacker 6 	 & 0.57 	 &  0.00 	 &  0.53 \\ 
\end{tabular}
\end{center}
\caption{Table of best scores for each attacker for each gesture. The results show that none of the attacks were successful: the passing score for the recognizer can be set so that
the Target can authenticate with ease and the attackers are not authenticated.}
\label{tab:ShoulderSurfingReg}
\end{table}

From the Table~\ref{tab:ShoulderSurfingReg} we see that there is not any overlapping for scores of target and attackers. This indicates our recognizer correctly differentiate attempts made by the target and that by attackers. In general, the target is always authenticating quite well across the the three gestures in comparison to the poorer scores of the attackers. The only opportunity where an attacker became close enough to steal the gesture (Attacker 2, Gesture 2) still has a one point cushion around it -- high enough to prevent authentication by an attacker if the threshold is set appropriately. 

\section{Discussion and Conclusions}

We have presented the first study of using free-form multitouch gestures
for mobile authentication. Towards the end of analyzing the security and memorability
of the gestures, we presented a novel metric based on computing the estimated
mutual information of repeated gestures. We designed and implemented a practical
multitouch recognizer as an authentication system, and studied the robustness of
free-form gestures against shoulder surfing attacks.

Overall, the results are favorable to user-generated free-form gestures as a means of authentication on touchscreen devices. 

Security, as estimated by $\hat{I}$, is high enough for most passwords the users generated. We learned that \emph{multi}finger gestures do not show high security in this measure. It should be noted though, most of multifinger gestures in our dataset are gestures of multiple fingers repeating the same simple shape, e.g.~drawing a circle with three fingers. We believe that the participants may overestimate the increase in security by merely increasing number of fingers; when they decided to use multiple fingers, they tended to choose a simple shape because they believed multiple fingers gave them high security despite simple shape. Such inconsistency in participants' perception and the actual security could be advised against in the password generation user interface. 

We also learned that, unlike with the length of a text password \cite{helpingusers}, the duration of a gesture does not play an important role in $\hat{I}$. Intuitively speaking, complex gestures with high $\hat{I}$ should take longer time to perform. However, we learned that even brief gestures can have high security. %
Gestures with duration less than two seconds have an average $\hat{I}$ less than 2\% lower than the average $\hat{I}$ for all gestures.

By looking at our dataset, we found out that some simple shapes, even circles, would actually take more time to complete than complex ones like signatures. The possible reasons warrant further studies. At this point, we suspect that complex gestures are also more difficult to reproduce precisely. A good secure gesture should have both: inherent complexity and easiness to perform. It is interesting in this light that signatures are particularly good and resulted in very high $\hat{I}$. This means, although very complex to perform, participants still managed to repeat them quite well. 

When it comes to memorability, the data show that users need a few repetitions to achieve a stable password. Like with text-based passwords, the generation of passwords is experienced as more demanding by users than recall. After generation, $\hat{I}$ drops after an interval of >10 days by 15.66 percent. However, they are still recognizable as unique passwords. In addition, for a participant, the value of $\hat{I}$ varies as they repeat the gesture. By continuously repeating, $\hat{I}$ tends to stabilize. Unlike the text-based passwords, which one has to input exactly, free-form gestures involve many sources of variance, which would be very difficult to keep constant across different attempts. Therefore, one alternative way of reporting security of gestures could be a range similar to confidence interval, instead of an exact value. Moreover, studying gesture variability is a good topic for future research, because a good balance must be found between memorability and security.

Several participants were able to create highly secure and memorable gestures. Below, we sketch strategies for generating such gestures. We plan to develop and test the guidelines and their effect on creating gestures in further studies.
These guidelines are illustrated with the best and worst gestures in Figure \ref{fig:gallerysummary}, especially with the worst gestures being simple multifinger circular motions.

\begin{enumerate}
\item General advice to promote consistency and retention: Practice different gestures first in order to get used to the touchscreen. Try out different gestures instead of picking the first one, to find one you prefer. Pick a gesture that will be used frequently to avoid forgetting it. Take more care and pay attention. Do not rush. Practice until faster and still accurate. Try to repeat each trace as closely as possible.
\item Characteristics of High $\hat{I}$ gestures to emulate: Use many sharp turns. Use a familiar gesture, for example, a signature. Use extra fingers to do different motions. Follow the above rules even when adding fingers.
\item Characteristics of Low $\hat{I}$ gestures to avoid: Do not use only few turns. Do not use gentle turns. Do not make turns to only to the same direction, for example, avoid doing a circle. 
\item Specific errors to avoid: Place fingers down in the same order each time. Use the same number of fingers each time.
\end{enumerate}

Our results from the recognizer show the capability of free-form multitouch gestures to work as passwords in a practical authentication system. The gestures were classified by the recognizer with relatively low error when being compared across all 63 participants,  indicating the ability of the participants to generate passwords that a recognizer would not have trouble classifying when comparing against multiple templates. The recognizer generated much higher scores when evaluating a participant's gesture against their template (ranging from 3 to 9) as compared to when it compared to other templates (ranging from 0 to 1). Memorability of the free-form gestures are also displayed through the EER values in recognizer results;  the lowest EER value for the first set is 3.34\% and the lowest for the second set is 13.16\%. The disparity in the data sets can be attributed to two factors: 1) the first data set had only two authentication trials compared to the second's five trials and 2) the second data set was performed after a much longer time span than in the first set, thus, there were memorability issues between the two sessions. The multitouch recognizer we designed and implemented has room for consideration in the future, for example, the effects of rotation, scale, and position invariance as added degrees of freedom with free-form multitouch gestures. %

Finally, with our preliminary shoulder surfing attack trial, we also learned that free-form gestures are relatively robust against shoulder surfing. None of the attackers were able to repeat the gestures well enough to be accepted by a practical authentication system. We acknowledge that further more comprehensive studies with several different kinds of gestures and more opportunities for the attackers would be warranted. For example, we could separate attackers into different groups in which half of them are allowed to rewatch the video recordings as many times they want to.

To conclude, our work shows that free-form gestures present a robust method of authentication for touchscreen devices.

\section*{Acknowledgments}

This material is based upon work supported by the National Science Foundation under Grant Number 1228777. Any opinions, findings, and conclusions or recommendations expressed in this material are those of the author(s) and do not necessarily reflect the views of the National Science Foundation.

\balance

\bibliographystyle{abbrv}
\bibliography{graphicalpasswords,incidental,securityadoption,janneown,privacy,usablesecurity,misc,pbw,passwords,passwordsblerta,smartphonesecurity,android,oulasvirta,infocapacity,gestures}

\begin{thebibliography}{10}

\bibitem{Aviv:2010:SAS:1925004.1925009}
A.~J. Aviv, K.~Gibson, E.~Mossop, M.~Blaze, and J.~M. Smith.
\newblock Smudge attacks on smartphone touch screens.
\newblock In {\em Proc. of WOOT'10}.

\bibitem{Biddle:2012:GPL:2333112.2333114}
R.~Biddle, S.~Chiasson, and P.~Van~Oorschot.
\newblock Graphical passwords: Learning from the first twelve years.
\newblock {\em ACM Comput. Surv.}, sep 2012.

\bibitem{Bo:2013:SSU:2500423.2504572}
C.~Bo, L.~Zhang, X.-Y. Li, Q.~Huang, and Y.~Wang.
\newblock Silentsense: Silent user identification via touch and movement
  behavioral biometrics.
\newblock In {\em Proc. of MobiCom '13}.

\bibitem{Bonneau:2012:SGA:2310656.2310721}
J.~Bonneau.
\newblock The science of guessing: Analyzing an anonymized corpus of 70 million
  passwords.
\newblock In {\em Proc. of SP '12}.

\bibitem{book}
Z.~Cai, C.~Shen, M.~Wang, Y.~Song, and J.~Wang.
\newblock Mobile authentication through touch-behavior features.
\newblock In {\em Biometric Recognition}, LNCS 8232. Springer International
  Publishing, 2013.

\bibitem{Chiasson:2009:MPI:1653662.1653722}
S.~Chiasson, A.~Forget, E.~Stobert, P.~C. van Oorschot, and R.~Biddle.
\newblock Multiple password interference in text passwords and click-based
  graphical passwords.
\newblock In {\em Proc. of CCS '09}.

\bibitem{cover-thomas}
T.~M. Cover and J.~A. Thomas.
\newblock {\em Elements of Information Theory}.
\newblock Wiley-Interscience, 2006.

\bibitem{DeLuca:2013:BAS:2470654.2481330}
A.~De~Luca, E.~von Zezschwitz, N.~D.~H. Nguyen, M.-E. Maurer, E.~Rubegni, M.~P.
  Scipioni, and M.~Langheinrich.
\newblock Back-of-device authentication on smartphones.
\newblock In {\em Proc. of CHI '13}.

\bibitem{Everitt:2009:CSF:1518701.1518837}
K.~M. Everitt, T.~Bragin, J.~Fogarty, and T.~Kohno.
\newblock A comprehensive study of frequency, interference, and training of
  multiple graphical passwords.
\newblock In {\em Proc. of CHI '09}.

\bibitem{hmm1}
J.~Fierrez, J.~Ortega-Garcia, D.~Ramos, and J.~Gonzalez-Rodriguez.
\newblock \uppercase{HMM}-based on-line signature verification: Feature
  extraction and signature modeling.
\newblock {\em Pattern Recogn. Lett.}, dec 2007.

\bibitem{Florencio:2007:LSW:1242572.1242661}
D.~Florencio and C.~Herley.
\newblock A large-scale study of web password habits.
\newblock In {\em Proc. of WWW'07}.

\bibitem{touchalytics}
M.~Frank, R.~Biedert, E.~Ma, I.~Martinovic, and D.~Song.
\newblock Touchalytics: On the applicability of touchscreen input as a
  behavioral biometric for continuous authentication.
\newblock {\em IEEE Trans. on Information Forensics and Security}, Jan 2013.

\bibitem{gogogate}
Gogogate.
\newblock \url{www.gogogate.com}.
\newblock Ref. Dec 3, 2013.

\bibitem{Grandhi:2011:UNI:1978942.1979061}
S.~A. Grandhi, G.~Joue, and I.~Mittelberg.
\newblock Understanding naturalness and intuitiveness in gesture production:
  insights for touchless gestural interfaces.
\newblock In {\em Proc. of CHI '11}.

\bibitem{hart:tlx}
S.~G. Hart and L.~E. Staveland.
\newblock {\em Development of NASA-TLX (Task Load Index): Results of empirical
  and theoretical research}.
\newblock 1988.

\bibitem{Jermyn:1999:DAG:1251421.1251422}
I.~Jermyn, A.~Mayer, F.~Monrose, M.~K. Reiter, and A.~D. Rubin.
\newblock The design and analysis of graphical passwords.
\newblock In {\em Proc. of SSYM'99}.

\bibitem{jones2006human}
L.~A. Jones and S.~J. Lederman.
\newblock {\em Human hand function}.
\newblock Oxford University Press, 2006.

\bibitem{Li:2010:PFA:1753326.1753654}
Y.~Li.
\newblock Protractor: a fast and accurate gesture recognizer.
\newblock In {\em Proc. of CHI '10}.

\bibitem{Long:2001:LSI:971478.971510}
A.~C. Long, J.~A. Landay, and L.~A. Rowe.
\newblock "those look similar!" issues in automating gesture design advice.
\newblock In {\em Proc. of PUI '01}.

\bibitem{azuremultifactor}
Microsoft.
\newblock Windows azure multi-factor authentication.
\newblock
  \url{www.windowsazure.com/en-us/documentation/services/multi-factor-authentication}.
  Ref. Dec 3, 2013.

\bibitem{hmm2}
D.~Muramatsu and T.~Matsumoto.
\newblock An \uppercase{HMM} on-line signature verifier incorporating signature
  trajectories.
\newblock In {\em Proc. of ICDAR '03}.

\bibitem{Oh:2013:CPE:2470654.2466145}
U.~Oh and L.~Findlater.
\newblock The challenges and potential of end-user gesture customization.
\newblock In {\em Proc. of CHI '13}.

\bibitem{Oorschot:2008:PMU:1284680.1284685}
P.~C.~v. Oorschot and J.~Thorpe.
\newblock On predictive models and user-drawn graphical passwords.
\newblock {\em ACM Trans. Inf. Syst. Secur.}, jan 2008.

\bibitem{oulasvirta2013information}
A.~Oulasvirta, T.~Roos, A.~Modig, and L.~Leppanen.
\newblock Information capacity of full-body movements.
\newblock In {\em Proc. of CHI'13}.

\bibitem{Pu:2013:WGR:2500423.2500436}
Q.~Pu, S.~Gupta, S.~Gollakota, and S.~Patel.
\newblock Whole-home gesture recognition using wireless signals.
\newblock In {\em Proc. of MobiCom '13}.

\bibitem{Rubine:1991:SGE:122718.122753}
D.~Rubine.
\newblock Specifying gestures by example.
\newblock In {\em Proc. of SIGGRAPH '91}.

\bibitem{Ruiz:2011:UMG:1978942.1978971}
J.~Ruiz, Y.~Li, and E.~Lank.
\newblock User-defined motion gestures for mobile interaction.
\newblock In {\em Proc. of CHI '11}.

\bibitem{Sae-Bae:2012:BGN:2207676.2208543}
N.~Sae-Bae, K.~Ahmed, K.~Isbister, and N.~Memon.
\newblock Biometric-rich gestures: a novel approach to authentication on
  multi-touch devices.
\newblock In {\em Proc. of CHI '12}.

\bibitem{Schaub:2013:EDS:2501604.2501615}
F.~Schaub, M.~Walch, B.~K\"{o}nings, and M.~Weber.
\newblock Exploring the design space of graphical passwords on smartphones.
\newblock In {\em Proc. of SOUPS '13}.

\bibitem{Serwadda:2013:KTB:2508859.2516659}
A.~Serwadda and V.~V. Phoha.
\newblock When kids' toys breach mobile phone security.
\newblock In {\em Proc. of CCS '13}.

\bibitem{square}
Square.
\newblock \url{www.squareup.com}.
\newblock Ref. Dec 3, 2013.

\bibitem{kinwrite}
J.~Tian, C.~Qu, W.~Xu, and S.~Wang.
\newblock Kinwrite: Handwriting-based authentication using kinect.
\newblock In {\em Proc. of NDSS '13}.

\bibitem{helpingusers}
B.~Ur, P.~Kelley, S.~Komanduri, J.~Lee, M.~Maass, M.~Mazurek, T.~Passaro,
  R.~Shay, T.~Vidas, L.~Bauer, N.~Christin, L.~Cranor, S.~Egelman, and
  J.~Lopez.
\newblock Helping users create better passwords.
\newblock {\em ;login}, Dec. 2012.

\bibitem{Wiedenbeck:2005:PDL:1090412.1090418}
S.~Wiedenbeck, J.~Waters, J.-C. Birget, A.~Brodskiy, and N.~Memon.
\newblock Passpoints: design and longitudinal evaluation of a graphical
  password system.
\newblock {\em Int. J. Hum.-Comput. Stud.}, 63(1-2):102--127, jul 2005.

\bibitem{Wobbrock:2007:GWL:1294211.1294238}
J.~O. Wobbrock, A.~D. Wilson, and Y.~Li.
\newblock Gestures without libraries, toolkits or training: a \$1 recognizer
  for user interface prototypes.
\newblock In {\em Proc. of UIST '07}.

\bibitem{Yan:2004:PMS:1024867.1025014}
J.~Yan, A.~Blackwell, R.~Anderson, and A.~Grant.
\newblock Password memorability and security: Empirical results.
\newblock {\em IEEE Security and Privacy}, sep 2004.

\bibitem{Zakaria:2011:SSD:2078827.2078835}
N.~H. Zakaria, D.~Griffiths, S.~Brostoff, and J.~Yan.
\newblock Shoulder surfing defence for recall-based graphical passwords.
\newblock In {\em Proc. of SOUPS '11}.

\bibitem{zhao:picturegesture}
Z.~Zhao and G.-J. Ahn.
\newblock On the security of picture gesture authentication.
\newblock In {\em Proc. of USENIX Security'13}.

\bibitem{WM-CS-2012-06}
N.~Zheng, K.~Bai, H.~Huang, and H.~Wang.
\newblock You are how you touch: User verication on smartphones via tapping
  behaviors.
\newblock Technical report, Dec. 2006.

\bibitem{zhou2009canonical}
F.~Zhou and F.~De~la Torre.
\newblock Canonical time warping for alignment of human behavior.
\newblock In {\em Proc. of NIPS'09}.

\end{thebibliography}
\end{document}